\documentclass[11pt,reqno]{article}
\usepackage{amssymb,amscd,amsmath,amsthm}
\def\proof{\noindent{\it Proof.}\enspace}
\def\endproof{\qed \medskip}
\newtheorem{thm}{Theorem}[section]
\newtheorem{prop}[thm]{Proposition}
\newtheorem{lemma}[thm]{Lemma}
\newtheorem{cor}[thm]{Corollary}

\newtheorem{remark}[thm]{Remark}

\def\cA{\mathcal{A}}
\def\ffi{\varphi}
\def\cB{\mathcal{B}}
\def\bC{\mathbb{C}}
\def\bZ{\mathbb{Z}}
\def\bR{\mathbb{R}}
\def\bN{\mathbb{N}}
\def\eps{\varepsilon}
\def\Tr{\mathrm{Tr}\,}
\def\<{\langle}
\def\>{\rangle}
\def\cH{\mathcal{H}}
\def\1{\mathbf{1}}
\def\ex{\mathrm{ex}\,}
\def\cE{\mathcal{E}}

\def\loc{\mathrm{loc}}
\def\sa{\mathrm{sa}}
\def\cS{\mathcal{S}}

\begin{document}

%\rightline{\today}
\ \vskip 1cm 
\centerline{\huge {\bf Free energy density for}} 
\bigskip
\centerline{\huge {\bf mean field perturbation of states}}
\bigskip
\centerline{\huge {\bf of a one-dimensional spin chain}}
\bigskip
\bigskip
\centerline{\large{\it Dedicated to Professor Walter Thirring on his 80th birthday}}
\bigskip
\bigskip
\centerline{\Large
Fumio Hiai\footnote{E-mail: hiai@math.is.tohoku.ac.jp; Partially supported by
Grant-in-Aid for Scientific Research (B)17340043.},
Mil\'an Mosonyi\footnote{E-mail: milan.mosonyi@gmail.com; Partially supported by
Grant-in-Aid for JSPS Fellows 18\,$\cdot$\,06916.},
Hiromichi Ohno\footnote{E-mail: ohno@math.kyushu-u.ac.jp; Partially supported by
Grant-in-Aid for JSPS Fellows 19\,$\cdot$\,2166.}
and D\'enes Petz\footnote{E-mail: petz@math.bme.hu; Partially supported by the
Hungarian Research Grant OTKA T068258.}}

\bigskip
\begin{center}
$^{1,\,2}$\,Graduate School of Information Sciences, Tohoku University \\
Aoba-ku, Sendai 980-8579, Japan
\end{center}
\begin{center}
$^3$\,Graduate School of Mathematics, Kyushu University, \\
1-10-6 Hakozaki, Fukuoka 812-8581, Japan
\end{center}
\begin{center}
$^4$\,Alfr\'ed R\'enyi Institute of Mathematics, \\ H-1364 Budapest,
POB 127, Hungary
\end{center}

\bigskip
\begin{abstract}
Motivated by recent developments on large deviations in states of the spin chain, we
reconsider the work of Petz, Raggio and Verbeure in 1989 on the variational expression
of free energy density in the presence of a mean field type perturbation. We extend
their results from the product state case to the Gibbs state case in the setting of
translation-invariant interactions of finite range. In the special case of a locally
faithful quantum Markov state, we clarify the relation between two different kinds of
free energy densities (or pressure functions).


\end{abstract}

\bigskip
\section{Introduction}

The theoretical description of the statistical mechanics of quantum spin chains was
the first success of the operator algebraic approach to quantum physics. A
one-dimensional spin chain is described by a quasi-local C*-algebra
$\cA:=\bigotimes_{k\in\bZ}\cA_k$  which is the infinite tensor product of full
matrix algebras $\cA_k=M_d(\bC)$  and the limit of the local algebras
$\cA_\Lambda:=\bigotimes_{k\in\Lambda}\cA_k$, where $\Lambda \subset \bZ$ is finite.
A state $\ffi$ of the spin chain is uniquely specified by its local restrictions
$\ffi_{\Lambda}:=\ffi|_{\cA_{\Lambda}}$. A local state $\omega$ of $\cA_{\Lambda}$
can equivalently be given by its density matrix $D(\omega)$ satisfying
$\omega(A)=\Tr D(\omega)A$, $A\in\cA_{\Lambda}$.

A translation-invariant interaction $\Phi$ of the spins determines a {\it local
Hamiltonian}
\begin{equation}\label{F-1.1}
H_\Lambda(\Phi):=\sum_{X\subset\Lambda}\Phi(X)
\end{equation}
with corresponding {\it local Gibbs state}
\begin{equation}\label{F-1.2}
\quad D(\ffi_\Lambda^G):=\frac{e^{-H_\Lambda(\Phi)}}{\Tr e^{-H_\Lambda(\Phi)}}
\end{equation}
for all finite $\Lambda\subset\bZ$. The local Gibbs state is the unique maximizer of
the functional $\omega\mapsto -\omega(H_{\Lambda}(\Phi))+S(\omega)$, where $\omega$
is an arbitrary state of $\cA_\Lambda$ and $S(\omega)$ is the von Neumann entropy
$S(\omega):=-\Tr D(\omega)\log D(\omega)$. Furthermore,
\begin{equation}\label{F-1.3}
\log\Tr_{\Lambda}e^{-H_{\Lambda}(\Phi)}=\max\{-\omega(H_{\Lambda}(\Phi))+S(\omega):
\omega \text{ state of } \cA_{\Lambda}\}\,.
\end{equation}
One of the main problems in the statistical mechanics of the spin chain is the
determination of the global equilibrium states of $\cA$ for a given interaction. When
$\Phi$ is of relatively short range, it is well known \cite{BR2,Is} that the
variational formula \eqref{F-1.3} holds in the asymptotic limit:
\begin{equation}\label{F-1.4}
P(\Phi) =\max\{-\omega(A_\Phi)+s(\omega):
\omega \text{ translation-invariant state of }\cA\}\,,
\end{equation}
where 
\begin{align}
P(\Phi)&:=\lim_{\Lambda\to\bZ}{1\over |\Lambda|}\log\Tr e^{-H_\Lambda(\Phi)}\,,
\label{F-1.5}\\
s(\omega)&:=\lim_{\Lambda\to\bZ}{1\over|\Lambda|}S(\omega|_{\cA_\Lambda})\,,
\label{F-1.6}\\
A_\Phi&:=\sum_{X\ni0}{\Phi(X)\over|X|}
\label{F-1.7}
\end{align}
are the {\it pressure} (or {\it free energy density}\,) of $\Phi$, the
{\it mean entropy} of $\omega$ and the {\it mean energy} of $\Phi$, respectively.
(Here note that the term ``free energy" should be used with minus sign in the exact
sense of physics.) Maximizers of the right-hand side of \eqref{F-1.4} are the
equilibrium states for the interaction $\Phi$. If $\Phi$ is of finite range, then the
equilibrium state is unique.

One of the main subjects of the present paper is an extension of the free energy
density \eqref{F-1.5} when the interaction is perturbed by a mean field term. Let
$\gamma$ be the right-translation automorphism of $\cA$ and set
$$
s_n(A):={1\over n}\sum_{\Lambda+k\subset[1,n]}\gamma^k(A)\in\cA_{[1,n]}
$$
for a fixed $A\in\cA_\Lambda^{\sa}$ with a finite $\Lambda\subset\bZ$. We will study
the limit
\begin{equation}\label{F-1.8}
\lim_{n \to \infty}{1\over n}\log\Tr \exp \bigl(-H_{[1,n]}(\Phi)-nf(s_n(A))\bigr)\,,
\end{equation}
where $f$ is a real continuous function. This kind of problem was initiated by Petz,
Raggio and Verbeure \cite{PRV} in the particular case when there is no interaction
between the spins. The motivation came from mean field models and the extension of
large deviation theory for quantum chains \cite{DP}. An important tool was St\o rmer's
quantum version of the de Finetti theorem for symmetric states. The subject was
treated in details in the monograph \cite{OP} under the name ``perturbational limits"
by using the concept of approximately symmetric sequences \cite{RW}. 
 Since the interaction $\Phi$ in the general situation
is not invariant under the permutation of the spins, our method in the general case
is the extremal decomposition theory for translation-invariant states that is standard
in quantum statistical mechanics, see \cite{BR1}. In the present paper we will show
that the limit is expressed by a variational formula generalizing \eqref{F-1.4}.

The limit \eqref{F-1.8} has a direct physical meaning in the case when $f(x)=x^2$ and
$A=A_0 \in \cA_0$. Then 
$$
-H_{[1,n]}(\Phi)-{1\over n}\sum_{i,j=1}^n A_iA_j
$$
is a mean field perturbation of the interaction $\Phi$, where $A_j:= \gamma^j (A_0)$.
The limit is the free energy density for the mean field model and the variational
formula has an important physical interpretation.

The limit density \eqref{F-1.8} can be considered in a different way as well. Given a
translation-invariant state $\ffi$, we can study the limit 
\begin{equation}\label{F-1.9}
p_\ffi(A,f):=\lim_{n\to\infty}{1\over n}\log
\Tr\exp \bigl(\log D(\ffi|_{\cA_{[1,n]}})-nf(s_n(A))\bigr)
\end{equation}
and its variational expression under the duality between the observable space
$\cA^\sa$ and the translation-invariant state space $\cS_\gamma(\cA)$. In particular,
when $f(x)=x$, the limit \eqref{F-1.9} becomes a simply perturbed free energy density
function (or pressure function)
$$
p_\ffi(A):=\lim_{n\to\infty}{1\over n}\log
\Tr\exp\bigl(\log D(\ffi|_{\cA_{[1,n]}})-ns_n(A)\bigr)
$$
for local observables $A$ in $\cA^{\sa}$ (if the limit exists). The dual
function of the function $p_\ffi(A)$ is the {\it mean relative entropy}
\begin{equation}\label{F-1.10}
S_M(\omega,\ffi)
:=\lim_{n\to\infty}{1\over n}S(\omega|_{\cA_{[1,n]}},\ffi|_{\cA_{[1,n]}})
\end{equation}
with respect to $\ffi$ defined for $\omega\in\cS_\gamma(\cA)$. The existence of the
mean relative entropy and its properties were worked out in \cite{HP0,HP2,HP3}.

When $\Phi$ is a translation-invariant interaction of finite range and $\ffi$ is
the equilibrium state for $\Phi$, the limits \eqref{F-1.8} and \eqref{F-1.9} are
the same (up to an additive term $P(\Phi)$), but \eqref{F-1.9} can also be studied
for a wider class of translation-invariant states, for example, for finitely
correlated states which were introduced by Fannes, Nachtergaele and Werner \cite{FNW}.
A slightly different concept of quantum Markov states was formerly introduced by
Accardi and Frigerio \cite{AF}. A translation-invariant and locally faithful quantum
Markov state in the sense of Accardi and Frigerio is known to be a finitely
correlated state as well as the equilibrium state for a nearest-neighbor interaction
\cite{AL,Oh}. Remarkably, a Markovian structure similar to the special quantum Markov
state just mentioned appears in the recent characterization \cite{HJPW,MP} of the
quantum states which saturate the strong subadditivity of von Neumann entropy.

A similar but different version of the free energy density function $p_\ffi(A)$ is
$$
\tilde p_\ffi(A):=\lim_{n\to\infty}{1\over n}\log
\ffi\bigl(e^{ns_n(A)}\bigr)
=\lim_{n\to\infty}{1\over n}\log
\ffi\Biggl(\exp\Biggl(\sum_{k=1}^n\gamma^k(A)\Biggr)\Biggr)\,,
$$
which gives the logarithmic moment generating function for a sequence of compactly
supported probability measures on the real line. Large deviations governed by this
generating function have recently been studied in \cite{NR,LR,HMO} for example.
In fact, our first motivation of the present paper came from large deviation results
in \cite{NR,LR} with respect to Gibbs-KMS states. It is not known in general for
$p_\ffi$ to have the interpretation as the logarithmic moment generating function as
$\tilde p_\ffi$ does. Indeed, this question is nothing more than the so-called
BMV-conjecture \cite{BMV}. On the other hand, since $\tilde p_\ffi$ is not a convex
function in general, it is impossible for $\tilde p_\ffi$ to enjoy such a variational
expression as $p_\ffi$ does.

The paper is organized as follows. Section 2 is a preliminary on translation-invariant
interactions and Gibbs-KMS equilibrium states of the one-dimensional spin chain. In
Section 3 the existence of the functional free energy density \eqref{F-1.9} and its
variational expression are obtained when $\ffi$ is the Gibbs state for a
translation-invariant interaction of finite range. In Section 4 the existence of
the density $p_\ffi(A)$ is proven for a general finitely correlated state $\ffi$,
and the exact relation between the functionals $p_\ffi$ and $\tilde p_\ffi$
introduced above is clarified in the special case when $\ffi$ is a locally faithful
quantum Markov state. Section 5 is a brief guide to how our results for a Gibbs state
$\ffi$ can be extended to the case of arbitrary dimension.
 
\section{Preliminaries}
\setcounter{equation}{0}

A one-dimensional spin chain is described by the infinite tensor product $C^*$-algebra
$\cA:=\bigotimes_{k\in\bZ}\cA_k$ of full matrix algebras $\cA_k:=M_d(\bC)$ over $\bZ$.
The right-translation automorphism of $\cA$ is denoted by $\gamma$. We denote by
$\cS_\gamma(\cA)$ the set of all $\gamma$-invariant states of $\cA$. The
$C^*$-subalgebra of $\cA$ corresponding to a subset $X$ of $\bZ$ is
$\cA_X:=\bigotimes_{k\in X}\cA_k$ with convention $\cA_\emptyset:=\bC\1$, where
$\1$ is the identity of $\cA$. If $X\subset Y\subset\bZ$, then $\cA_X\subset\cA_Y$
by a natural inclusion. The local algebra is the dense $*$-subalgebra
$\cA_{\loc}:=\bigcup_{n=1}^\infty\cA_{[-n,n]}$ of $\cA$. The self-adjoint parts of
$\cA_{\loc}$ and $\cA$ are denoted by $\cA_{\loc}^{\sa}$ and $\cA^{\sa}$, respectively.
The usual trace on $\cA_X$ for each finite $X\subset\bZ$ is denoted by $\Tr$ without
referring to $X$ since it causes no confusion.

An interaction $\Phi$ in $\cA$ is a mapping from the nonempty finite subsets of $\bZ$
into $\cA$ such that $\Phi(X)=\Phi(X)^*\in\cA_X$ for each finite $X\subset\bZ$. Given
an interaction $\Phi$ and a finite subset $\Lambda\subset\bZ$, we have the local
Hamiltonian $H_\Lambda(\Phi)$ given in \eqref{F-1.1} and the {\it surface energy}
$W_\Lambda(\Phi)$
$$
W_\Lambda(\Phi):=\sum\{\Phi(X):
X\cap\Lambda\ne\emptyset,\,X\cap\Lambda^c\ne\emptyset\}
$$
whenever the sum converges in norm. We always assume that $\Phi$ is $\gamma$-invariant,
i.e., $\gamma(\Phi(X))=\Phi(X+1)$ for every finite $X\subset\bZ$, where
$X+1:=\{k+1:k\in X\}$. We denote by $\cB_0(\cA)$ the set of all $\gamma$-invariant
interactions $\Phi$ in $\cA$ such that
$$
\|\Phi\|_0:=\sum_{X\ni0}\|\Phi(X)\|+\sup_{n\ge1}\|W_{[1,n]}(\Phi)\|<+\infty\,.
$$
It is easy to see that $\cB_0(\cA)$ is a real Banach space with the usual linear
operations and the norm $\|\Phi\|_0$. Associated with $\Phi\in\cB_0(\cA)$ we have a
strongly continuous one-parameter automorphism group $\alpha^\Phi$ of $\cA$ given by
$$
\alpha^\Phi_t(A)=\lim_{m\to-\infty,n\to\infty}
e^{itH_{[m,n]}(\Phi)}Ae^{-itH_{[m,n]}(\Phi)}
\qquad (A\in\cA)\,.
$$
Then it is known \cite{Ar1,Ki} that there exists a unique {\it $\alpha^\Phi$-KMS
state} (at $\beta=-1$) $\ffi$ of $\cA$, which is automatically faithful and ergodic 
(i.e., an extremal point of $\cS_\gamma(\cA)$). The KMS state $\ffi$ is
characterized by the Gibbs condition and so it is also called the (global)
{\it Gibbs state} for $\Phi$. The state $\ffi$ is also characterized by the
{\it variational principle} $s(\ffi)=\ffi(A_\Phi)+P(\Phi)$, the equality case of the
expression \eqref{F-1.4}, where $P(\Phi)$, $s(\ffi)$ and $A_\Phi$ are given in
\eqref{F-1.5}--\eqref{F-1.7}. See \cite{BR2,Is} for details on these equivalent
characterizations of equilibrium states.

In the rest of this section, assume that $\Phi$ is a $\gamma$-invariant interaction
of finite range, i.e., there is an $N_0\in\bN$ such that $\Phi(X)=0$ whenever the
diameter of $X$ is greater than $N_0$. Of course, $\Phi\in\cB_0(\cA)$. Let $\ffi$ be
the $\alpha^\Phi$-KMS state (at $\beta=-1$) of $\cA$. The next lemma will play an
essential role in our discussions below; the proof can be found in \cite{Ar0,Ar2,AI}.

\begin{lemma}\label{L-2.1}
There is a constant $\lambda\ge1$ {\rm(}independent of $n${\rm)} such that
$$
\lambda^{-1}\ffi_n\le\ffi_n^G\le\lambda\ffi_n
$$
for all $n\in\bN$, where $\ffi_n^G$ is the local Gibbs state \eqref{F-1.2} with
$\Lambda=[1,n]$.
\end{lemma}

For $\omega\in\cS_\gamma(\cA)$ and $\Psi\in\cB_0(\cA)$ we write for short $\omega_n$
and $H_n(\Psi)$ for $\omega|_{\cA_{[1,n]}}$ and $H_{[1,n]}(\Psi)$, respectively.
Lemma \ref{L-2.1} gives
$$
\bigg|{1\over n}\log\Tr\exp\bigl(\log D(\ffi_n)-H_n(\Psi)\bigr)
-{1\over n}\log\Tr\exp\bigl(\log D(\ffi_n^G)-H_n(\Psi)\bigr)\bigg|
\le{\log\lambda\over n}\,.
$$
Since
$$
\Tr\exp\bigl(\log D(\ffi_n^G)-H_n(\Psi)\bigr)
={\Tr e^{-H_n(\Phi+\Psi)}\over\Tr e^{-H_n(\Phi)}}\,,
$$
we have

\begin{lemma}\label{L-2.2}
For every $\Psi\in\cB_0(\cA)$ the limit
$$
P_\ffi(\Psi):=\lim_{n\to\infty}{1\over n}
\log\Tr\exp\bigl(\log D(\ffi_n)-H_n(\Psi)\bigr)
$$
exists and
$$
P_\ffi(\Psi)=P(\Phi+\Psi)-P(\Phi)\,.
$$
\end{lemma}

For every $\omega\in\cS_\gamma(\cA)$ the mean relative entropy \eqref{F-1.10} exists
and
\begin{equation}\label{F-2.1}
S_M(\omega,\ffi)
=\lim_{n\to\infty}{1\over n}S(\omega_n,\ffi_n)
=\lim_{n\to\infty}{1\over n}S(\omega_n,\ffi_n^G)\,,
\end{equation}
see \cite[p.\,710]{HP2}. In fact, since
$$
S(\omega_n,\ffi_n^G)
=-S(\omega_n)+\omega(H_n(\Phi))+\log\Tr e^{-H_n(\Phi)}
$$
and
$$
\lim_{n\to\infty}{\omega(H_n(\Phi))\over n}=\omega(A_\Phi)\,,
$$
we have

\begin{lemma}\label{L-2.3}
For every $\omega\in\cS_\gamma(\cA)$,
$$
S_M(\omega,\ffi)=-s(\omega)+\omega(A_\Phi)+P(\Phi)\,.
$$
Hence, the function $\omega\mapsto S_M(\omega,\ffi)$ is affine and lower
semicontinuous in the weak* topology on $\cS_\gamma(\cA)$.
\end{lemma}

\begin{thm}\label{T-2.4} \
\begin{itemize}
\item[\rm(a)] For every $\Psi\in\cB_0(\cA)$,
$$
P_\ffi(\Psi)=\max\{-\omega(A_\Psi)-S_M(\omega,\ffi):\omega\in\cS_\gamma(\cA)\}\,.
$$
\item[\rm(b)] For every $\omega\in\cS_\gamma(\cA)$,
$$
S_M(\omega,\ffi)=\sup\{-\omega(A_\Psi)-P_\ffi(\Psi):\Psi\in\cB_0(\cA)\}\,.
$$
\item[\rm(c)] The function $P_\ffi$ on $\cB_0(\cA)$ is G\^ateaux-differentiable at any
$\Psi\in\cB_0(\cA)$, i.e., the limit
$$
\partial(P_\ffi)_\Psi(\Psi')
:=\lim_{t\to0}{P_\ffi(\Psi+t\Psi')-P_\ffi(\Psi)\over t}
$$
exists for every $\Psi'\in\cB_0(\cA)$. Moreover, when $\ffi^\Psi$ is the unique
$\alpha^{\Phi+\Psi}$-KMS state,
$$
\partial(P_\ffi)_\Psi(\Psi')=-\ffi^\Psi(A_{\Psi'}).
$$
\end{itemize}
\end{thm}

\proof 
The variational expressions in (a) and (b) are just rewriting of \eqref{F-1.4} and
$$
s(\omega)=\inf\{\omega(A_\Psi)+P(\Psi):\Psi\in\cB_0(\cA)\}
$$
thanks to Lemmas \ref{L-2.2} and \ref{L-2.3} (see \cite[\S II.3]{Is} for the above
expression of $s(\omega)$ complementary to \eqref{F-1.4}). Note also that the maximum
in (a) is attained by the unique Gibbs state for $\Phi+\Psi$.

The differentiability of $P_\ffi$ in (c) was essentially shown in
\cite[Corollary 3.5]{LR}; we give the proof for completeness. Let $\cB_0(\cA)^*$ be
the dual Banach space of $\cB_0(\cA)$. For each $\omega\in\cS_\gamma(\cA)$ define
$f_\omega\in\cB_0(\cA)^*$ by $f_\omega(\Psi):=-\omega(A_\Psi)$. Then
$\omega\mapsto f_{\omega}$ is an injective and continuous (in the weak* topologies)
affine map \cite[Lemma II.1.1]{Is}; hence 
$\Gamma:=\{f_\omega:\omega\in\cS_\gamma(\cA)\}$ is a weak* compact convex subset of
$\cB_0(\cA)^*$ and 
$$
F(f):=\begin{cases}
S_M(\omega,\ffi) & \text{if $f=f_\omega$ with $\omega\in\cS_\gamma(\cA)$}, \\
+\infty & \text{otherwise}
\end{cases}
$$
is a well-defined function on $\cB_0(\cA)^*$ which is convex and weakly* lower
semicontinuous. The assertion (a) means that $P_\ffi$ is the conjugate function of
$F$, which in turn implies that the conjugate function of $P_\ffi$ on $\cB_0(\cA)$
is $F$. By the general theory of conjugate functions (see
\cite[Proposition I.5.3]{ET} for example), $P_\ffi$ is G\^ateaux-differentiable at
$\Psi\in\cB_0(\cA)$ if and only if there is a unique $f\in\cB_0(\cA)^*$ such that
$(P_\ffi)^*(f)=f(\Psi)-P_\ffi(\Psi)$, that is, there is a unique
$\ffi^\Psi\in\cS_\gamma(\cA)$ such that
\begin{equation}\label{F-2.2}
S_M(\ffi^\Psi,\ffi)=-\ffi^\Psi(A_\Psi)-P_\ffi(\Psi)\,.
\end{equation}
By Lemmas \ref{L-2.2} and \ref{L-2.3} the above equality is equivalent to the
variational principle
$$
s(\ffi^\Psi)=\ffi^\Psi(A_{\Phi+\Psi})+P(\Phi+\Psi)\,,
$$
which is equivalent to $\ffi^\Psi$ being the $\alpha^{\Phi+\Psi}$-KMS state. Hence
the differentiability assertion of $P_\ffi$ follows. Moreover, by (a) we get
$$
P_\ffi(\Psi+t\Psi')\ge-\ffi^\Psi(A_{\Psi+t\Psi'})-S_M(\omega,\ffi)
$$
for any $\Psi'\in\cB_0(\cA)$ and $t\in\bR$. This together with equality \eqref{F-2.2}
for $t=0$ gives the formula $\partial(P_\ffi)_\Psi(\Psi')=-\ffi^\Psi(A_{\Psi'})$.
\endproof 

\begin{cor}\label{C-2.5} \
\begin{itemize}
\item[\rm(1)] For every $A\in\cA_{\loc}^{\sa}$ so that $A\in\cA_\Lambda^{\sa}$ with a
finite $\Lambda\subset\bZ$, the free energy density
\begin{equation}\label{F-2.3}
p_\ffi(A):=\lim_{n\to\infty}{1\over n}
\log\Tr\exp\Biggl(\log D(\ffi_n)
-\sum_{\Lambda+k\subset[1,n]}\gamma^k(A)\Biggr)
\end{equation}
exists {\rm(}independently of the choice of $\Lambda${\rm)}.
\item[\rm(2)] The function $p_\ffi$ on $\cA_{\loc}^{\sa}$ is G\^ateaux-differentiable
at any $A\in\cA_{\loc}^{\sa}$ in the sense that the limit
$$
\lim_{t\to0}{p_\ffi(A+tB)-p_\ffi(A)\over t}
$$
exists for every $B\in\cA_{\loc}^{\sa}$. In particular, the function
$t\in\bR\mapsto p_\ffi(tA)$ is differentiable for every $A\in\cA_{\loc}^{\sa}$.
\item[\rm(3)] The above function $p_\ffi$ on $\cA_{\loc}^{\sa}$ uniquely extends to a
function {\rm(}denoted by the same $p_\ffi${\rm)} on $\cA^{\sa}$ which is convex and
Lipschitz continuous with
$$
|p_\ffi(A)-p_\ffi(B)|\le\|A-B\|\,,\qquad A,B\in\cA^{\sa}\,.
$$
\item[\rm(4)] For every $A\in\cA^{\sa}$,
$$
p_\ffi(A)=\max\{-\omega(A)-S_M(\omega,\ffi):\omega\in\cS_\gamma(\cA)\}.
$$
\item[\rm(5)] For every $\omega\in\cS_\gamma(\cA)$,
\begin{align*}
S_M(\omega,\ffi)&=\sup\{-\omega(A)-p_\ffi(A):A\in\cA_{\loc}^{\sa}\} \\
&=\sup\{-\omega(A)-p_\ffi(A):A\in\cA^{\sa}\}\,.
\end{align*}
\end{itemize}
\end{cor}

\proof 
To show (1), we may assume $A\in\cA_{[1,\ell(A)]}^{\sa}$ with some $\ell(A)\in\bN$,
and set a $\gamma$-invariant interaction $\Psi_A$ of finite range (hence
$\Psi_A\in\cB_0(\cA)$) by
$$
\Psi_A(X):=\begin{cases}
\gamma^k(A) & \text{if $X=[k+1,k+\ell(A)]$, $k\in\bZ$}, \\
0 & \text{otherwise}.
\end{cases}
$$
Since $\sum_{k=0}^{n-\ell(A)}\gamma^k(A)=H_n(\Psi_A)$, the limit \eqref{F-2.3} exists
by Lemma \ref{L-2.2} and its independence of the choice of $\Lambda$ is obvious.
The differentiability in (2) immediately follows from Theorem \ref{T-2.4}\,(c).
(In fact, the derivative of $p_\ffi$ at $A$ is $\partial(p_\ffi)_A(B)=-\ffi^A(B)$ for
every $B\in\cA_\loc^\sa$, where $\ffi^A$ is the unique $\alpha^{\Phi+\Psi_A}$-KMS
state.) Moreover, since
$$
A_{\Psi_A}={1\over\ell(A)}\sum_{k=1}^{\ell(A)}\gamma^{-k}(A)
$$
so that $\omega(A_{\Psi_A})=\omega(A)$ for all $\omega\in\cS_\gamma(\cA)$, Theorem
\ref{T-2.4}\,(a) implies the variational expression in (4) for any
$A\in\cA_{\loc}^{\sa}$. The Lipschitz inequality in (3) for every
$A,B\in\cA_{\loc}^{\sa}$ is immediately seen from the formula \eqref{F-2.3}. Hence
$p_\ffi$ uniquely extends to a Lipschitz continuous function on $\cA^{\sa}$, and the
convexity of $p_\ffi$ on $\cA^{\sa}$ is obvious. To prove (4) for general
$A\in\cA^{sa}$ let $\{A_n\}$ be a sequence in $\cA_\loc^\sa$ such that $\|A_n-A\|\to0$.
It is clear by convergence that $p_\ffi(A)\ge-\omega(A)-S_M(\omega,\ffi)$ for all
$\omega\in\cS_\gamma(\cA)$. Let $\omega_n$ be the maximizer of the right-hand side of
(4) for $A_n$; here it may be assumed that $\{\omega_n\}$ converges to
$\omega\in\cS_\gamma(\cA)$ in the weak* topology. Then we get
$$
p_\ffi(A)=\lim_{n\to\infty}\{-\omega_n(A_n)-S_M(\omega_n,\ffi)\}
\le\omega(A)-S_M(\omega,\ffi)
$$
by Lemma \ref{L-2.3} (the weak* lower semicontinuity), which proves (4). Finally, (5)
follows from Lemma \ref{L-2.3} and the duality theorem for conjugate functions
\cite[Proposition I.4.1]{ET}.
\endproof

For each $A\in\cA^{\sa}$ we have the convex and continuous function
$t\mapsto p_\ffi(tA)$ on $\bR$ by Corollary \ref{C-2.5}\,(3). We now introduce the
function
\begin{equation}\label{F-2.4}
I_A(x):=\inf\{S_M(\omega,\ffi):\omega\in\cS_\gamma(\cA),\,\omega(A)=x\}
\qquad (x\in\bR)\,.
\end{equation}
Obviously, $I_A(x)=+\infty$ for $x\not\in[\lambda_{\min}(A),\lambda_{\max}(A)]$,
where $\lambda_{\min}(A)$ and $\lambda_{\max}(A)$ are the minimum and the maximum of
the spectrum of $A$. The next proposition says that $p_\ffi(tA)$ and $I_A(x)$ are
the Legendre transforms of each other, which are the contractions of the expressions
in the above (5) and (4) into the real line via $\omega\mapsto\omega(A)$.

\begin{prop}\label{P-2.6}
For every $A\in\cA^{\sa}$,
\begin{align*}
I_A(x)&=\sup\{-tx-p_\ffi(tA):t\in\bR\}\,,\qquad x\in\bR\,, \\
p_\ffi(tA)&=\max\{-tx-I_A(x):x\in[\lambda_{\min}(A),\lambda_{\max}(A)]\}\,,
\qquad t\in\bR\,.
\end{align*}
\end{prop}

\proof 
We have
\begin{align*}
I_A(x)&=\min_{\omega\in\cS_\gamma(\cA)}\sup_{t\in\bR}
\{t(-x+\omega(A))+S_M(\omega,\ffi)\} \\
&=\sup_{t\in\bR}\min_{\omega\in\cS_\gamma(\cA)}
\{t(-x+\omega(A))+S_M(\omega,\ffi)\} \\
&=\sup_{t\in\bR}\{-tx-p_\ffi(tA)\}
\end{align*}
by Corollary \ref{C-2.5}\,(4). In the above, the second equality follows from Sion's
minimax theorem \cite{Si} thanks to Lemma \ref{L-2.3}. (The elementary proof in
\cite{Ko} for real-valued functions can also work for functions with values in
$(-\infty,+\infty]$.) The second formula follows from the first by duality.
\endproof 

\begin{remark}\label{R-2.7}{\rm
An alternative notion of free energy density  
\begin{equation}\label{F-2.5}
\tilde p_\ffi(A):=\lim_{n\to\infty}{1\over n}
\log\ffi\Biggl(\exp\Biggl(\sum_{k=0}^{n-1}\gamma^k(A)\Biggr)\Biggr)
\end{equation}
was recently studied in \cite{NR,LR,HMO} in relation with large deviation problems on
the spin chain. The function $t\in\bR\mapsto \tilde p_\ffi(tA)$ is the so-called
logarithmic moment generating function \cite{DZ} of a sequence of probability measures
and existence of the limit guarantees large deviation upper bound to hold, while if
the limit is even differentiable that provides full large deviation principle. The
existence of the limit was proven for any $A\in\cA_{\loc}^{\sa}$ when $\ffi$ is the
unique Gibbs state of a translation-invariant interaction of finite range \cite{LR}
and when $\ffi$ is a finitely correlated state \cite{HMO}. Differentiability was shown
in \cite{NR} and \cite{HMO} for certain special cases. The Golden-Thompson inequality
shows that
\begin{equation}\label{F-2.6}
p_\ffi(A)\le\tilde p_\ffi(A)
\end{equation}
holds for any $A\in\cA_{\loc}^{\sa}$. For instance, for a product state
$\ffi=\bigotimes_\bZ\rho$ with $D(\rho)=e^{-H}$ and a one-site observable $A$, since
$\tilde p_\ffi(A)=\log\Tr(e^{-H}e^{-A})$ while $p_\ffi(A)=\log\Tr(e^{-H-A})$, the
equality $p_\ffi(A)=\tilde p_\ffi(A)$ occurs only when $A$ commutes with $H$ (see
\cite{Hi}). Although the Lipschitz continuity of $\tilde p_\ffi$ on $\cA_{\loc}^{\sa}$
and its variational expression as in the above (4) are impossible, it might be possible
to get the variational expression as in (5) with $\tilde p_\ffi$ in place of $p_\ffi$.
This is equivalent to saying that $p_\ffi$ on $\cA^{\sa}$ is the lower semicontinuous
convex envelope of $\tilde p_\ffi$ on $\cA_{\loc}^{\sa}$, as will be shown in a
special case in Section 4 (see Corollary \ref{C-4.10}).
}\end{remark}

\begin{remark}\label{R-2.8}{\rm
An equivalent formulation of the celebrated conjecture due to Bessis, Moussa and
Villani \cite{BMV} (the so-called BMV-conjecture) is stated as follows \cite{LS}:
If $H_0$ and $H_1$ are $N\times N$ Hermitian matrices with $H_1\ge0$, then there
exists a positive measure $\mu$ on $[0,\infty)$ such that
$$
\Tr\,e^{H_0-tH_1}=\int_0^\infty e^{-ts}\,d\mu(s)\,,\qquad t>0\,;
$$
or equivalently, the function $\Tr\,e^{H_0-tH_1}$ on $t>0$ is completely monotone. Now
if the BMV-conjecture held true with
$$
H_0:=\log D(\ffi_n)\,,\quad
H_1:={1\over n}\sum_{\Lambda+k\subset[1,n]}\gamma^k(A)\,,
$$
where $A\in\cA_\Lambda^{\sa}$ with a finite $\Lambda\subset\bZ$, we would have a
probability measure $\mu_n$ supported in $[\lambda_{\min}(A),\lambda_{\max}(A)]$ such
that
$$
\Tr\exp\Biggl(\log D(\ffi_n)
-\sum_{\Lambda+k\subset[1,n]}\gamma^k(tA)\Biggr)
=\int_{-\infty}^\infty e^{-nts}\,d\mu_n(s)\,,\qquad t\in\bR\,.
$$
(The restriction on the support of $\mu_n$ easily follows from the Paley-Wiener
theorem.) In this situation, the free energy density $p_\ffi(tA)$ is the logarithmic
moment generating function of the sequence of measures $(\mu_n)$, and Corollary
\ref{C-2.5} and Proposition \ref{P-2.6} combined with the G\"artner-Ellis theorem
\cite[Theorem 2.3.6]{DZ} yield that $(\mu_n)$ satisfies the large deviation
principle with the good rate function $I_A(x)$ given in \eqref{F-2.4}.
}\end{remark}

\section{Perturbation of Gibbs states}
\setcounter{equation}{0}

When the reference state $\ffi$ is a product state and $A$ is a one-site observable,
the variational expression of functional free energy density
\begin{align*}
&\lim_{n\to\infty}{1\over n}
\log\Tr \exp \bigl(\log D(\ffi_n) -nf(s_n(A))\bigr) \\
&\qquad\qquad= \sup_{\omega}\Big\{-\lim_{n \to \infty}\omega(f(s_n(A)))
-S_M(\omega, \ffi)\Big\}
\end{align*}
was obtained in \cite{PRV}, where $\omega$ runs over the symmetric (or
permutation-invariant) states. A comprehensive exposition on the subject is also
found in \cite[\S13]{OP}, which contains a generalization of the above expression
though $\ffi$ is still a product state. In this section we consider the case when
the reference state $\ffi$ is the Gibbs state for a translation-invariant
interaction $\Phi$ of finite range.

Let $A\in\cA_{\loc}^{\sa}$. We may assume without loss of generality that
$A\in\cA_{[1,\ell(A)]}^{\sa}$ with some $\ell(A)\in\bN$, and set
$$
s_n(A):=\frac{1}{n}\sum_{k=0}^{n-\ell(A)}\gamma^k(A) \in \cA_{[1,n]}\,.
$$
Given $A$ and a continuous function $f:[\lambda_{\min}(A),\lambda_{\max}(A)]\to\bR$
the {\it functional free energy density} is defined as the limit
$$
\lim_{n\to\infty}{1\over n}\log Z_\ffi(n,A,f)
$$
for
$$
Z_\ffi(n,A,f):=\Tr\exp\bigl(\log D(\ffi_n)-nf(s_n(A))\bigr)
$$
as $n \to \infty$. We will show the existence of the limit in Theorem \ref{T-3.4}.

The extreme boundary $\ex\cS_\gamma(\cA)$ of the set $\cS_\gamma(\cA)$ consists of
the ergodic states. It is known that $\ex\cS_\gamma(\cA)$ is a $G_\delta$-subset of
$\cS_\gamma(\cA)$ (see \cite[Proposition 1.3]{Ph}). Since $(\cA,\gamma)$ is
asymptotically Abelian in the norm sense, $\cS_\gamma(\cA)$ is a so-called Choquet
simplex (see \cite[Corollary 4.3.11]{BR1}) so that each $\omega\in\cS_\gamma(\cA)$
has a unique extremal decomposition
$$
\omega=\int_{\ex\cS_\gamma(\cA)}\psi\,d\nu_\omega(\psi)
$$
with a probability Borel measure $\nu_\omega$ on $\ex\cS_\gamma(\cA)$ (see
\cite[p.\,66]{Ph}, \cite[Theorem 4.1.15]{BR1}).

\begin{lemma}\label{L-3.1}
For every continuous $f:[\lambda_{\min}(A),\lambda_{\max}(A)] \to \bR$ and for every
$\omega\in\cS_\gamma(\cA)$ the limit
$$
E_{A,f}(\omega):=\lim_{n\to\infty}\omega(f(s_n(A)))
$$
exists and
$$
E_{A,f}(\omega)=\int_{\ex\cS_\gamma(\cA)}f(\psi(A))\,d\nu_\omega(\psi)
$$
for the extremal decomposition
$\omega=\int_{\ex\cS_\gamma(\cA)}\psi\,d\nu_\omega(\psi)$.
\end{lemma}

\proof 
The first assertion is contained in \cite[Proposition 13.2]{OP}. However, we use a
different  method to prove the two statements together.

First let $\psi\in\ex\cS_\gamma(\cA)$ and
$(\pi_\psi,\cH_\psi,U_\psi,\Omega_\psi)$ be the GNS construction associated with
$\psi$, i.e., $\pi_\psi$ is a representation of $\cA$ on $\cH_\psi$ with a cyclic
vector $\Omega_\psi$ and $U_\psi$ is a unitary on $\cH_\psi$ such that
$\psi(A)=\<\pi_\psi(A)\Omega_\psi,\Omega_\psi\>$ and
$\pi_\psi(\gamma(A))=U_\psi\pi_\psi(A)U_\psi^*$ for all $A\in\cA$. Thanks to the
asymptotic Abelianness, the extremality of $\psi$ means (see
\cite[Theorem 4.3.17]{BR1}) that the set of $U_\psi$-invariant vectors in $\cH_\psi$
is the one-dimensional subspace $\bC\Omega_\psi$. Hence the mean ergodic theorem
implies that
$$
\pi_\psi(s_n(A))\Omega_\psi
={1\over n}\sum_{k=0}^{n-\ell(A)}U_\psi^k\pi_\psi(A)\Omega_\psi
$$
converges in norm to $\psi(A)\Omega_\psi$ as $n\to\infty$. The case $f(x)=x^m$
easily follows from this, and by approximating $f$ by polynomials, we get
$$
\lim_{n\to\infty}\|\pi_\psi(f(s_n(A)))\Omega_\psi-f(\psi(A))\Omega_\psi\|=0
$$
so that
$$
\lim_{n\to\infty}\psi(f(s_n(A)))=f(\psi(A))\,.
$$
Finally, for a general $\omega\in\cS_\gamma(\cA)$
with the extremal decomposition
$\omega=\int_{\ex\cS_\gamma(\cA)}\psi\,d\nu_\omega(\psi)$, the Lebesgue convergence
theorem gives
$$
\lim_{n\to\infty}\omega(f(s_n(A)))
=\lim_{n\to\infty}\int_{\ex\cS_\gamma(\cA)}
\psi(f(s_n(A)))\,d\nu_\omega(\psi)
=\int_{\ex\cS_\gamma(\cA)}f(\psi(A))\,d\nu_\omega(\psi)\,,
$$
as required.
\endproof 

In the following proofs we will often use a state perturbation technique. For the
convenience of the reader, we here summarize some basic properties of state
perturbation restricted to the simple case of matrix algebras. See \cite{BR2,OP} for
the general theory of the subject matter. Let $\rho$ be a faithful state of
$\cB:=M_N(\bC)$ with density matrix $e^{-H}$. For each $h\in\cB^{\sa}$  define the
perturbed functional $\rho^h$ by
$$
\rho^h(A):=\Tr e^{-H-h}A\qquad (A\in\cB)
$$
and the normalized version 
$$
[\rho^h](A):={\rho^h(A)\over\rho^h(\1)}
={\Tr e^{-H-h}A\over\Tr e^{-H-h}}\qquad (A\in\cB)\,.
$$
The state $[\rho^h]$ is characterized as the unique minimizer of the functional
$$
\omega\mapsto S(\omega,\rho)+\omega(h)
$$
on the states of $\cB$. It is plain to see the chain rule:
$[[\rho^h]^k]=[\rho^{h+k}]$ for all $h,k\in\cB^{\sa}$. For each state $\omega$ of
$\cB$, from the equality
$$
S(\omega,[\rho^h])=S(\omega,\rho)+\omega(h)+\log\rho^h(\1)
$$
and the Golden-Thompson inequality $\rho^h(\1)\le\rho(e^{-h})$, the following are
readily seen:
\begin{equation}\label{F-3.1}
\log\rho^h(\1)\ge-\omega(h)-S(\omega,\rho)\,,
\end{equation}
\begin{equation}\label{F-3.2}
|S(\omega,\rho)-S(\omega,[\rho^h])|\le2\|h\|\,.
\end{equation}

\begin{lemma}\label{L-3.2}
For every continuous $f:[\lambda_{\min}(A),\lambda_{\max}(A)] \to \bR$ and for every
$\omega\in\cS_\gamma(\cA)$,
$$
\liminf_{n\to\infty}{1\over n}\log Z_\ffi(n,A,f)
\ge\sup\{-E_{A,f}(\omega)-S_M(\omega,\ffi):\omega\in\cS_\gamma(\cA)\}
$$
holds.
\end{lemma}

\proof 
For $n\in\bN$ write $h_n:=nf(s_n(A))$ for simplicity. The perturbed functional
$\ffi_n^{h_n}$ of $\ffi_n$ on $\cA_{[1,n]}$ has the density $\exp(\log D(\ffi_n)-h_n)$
and so $Z_\ffi(n,A,f)=\ffi_n^{h_n}(\1)$. Hence it follows from \eqref{F-3.1} that
$$
\log Z_\ffi(n,A,f)\ge-\omega_n(h_n)-S(\omega_n,\ffi_n)\,,
\qquad\omega\in\cS_\gamma(\cA)\,.
$$
By Lemma \ref{L-3.1} and \eqref{F-2.1} we have
$$
\liminf_{n\to\infty}{1\over n}\log Z_\ffi(n,A,f)
\ge-E_{A,f}(\omega)-S_M(\omega,\ffi)
$$
for all $\omega\in\cS_\gamma(\cA)$.
\endproof 

\begin{lemma}\label{L-3.3}
For every continuous $f:[\lambda_{\min}(A),\lambda_{\max}(A)] \to \bR$,
\begin{align*}
&\sup\{-E_{A,f}(\omega)-S_M(\omega,\ffi):\omega\in\cS_\gamma(\cA)\} \\
&\qquad=\sup\{-f(\psi(A))-S_M(\psi,\ffi):\psi\in\ex\cS_\gamma(\cA)\} \\
&\qquad=\max\{-f(\omega(A))-S_M(\omega,\ffi):\omega\in\cS_\gamma(\cA)\} \\
&\qquad=\max\{-f(x)-I_A(x):x\in[\lambda_{\min}(A),\lambda_{\max}(A)]\}\,.
\end{align*}
\end{lemma}

\proof 
For every $\omega\in\cS_\gamma(\cA)$ let
$\omega=\int_{\ex\cS_\gamma(\cA)}\psi\,d\nu_\omega(\psi)$ be the extremal
decomposition of $\omega$. By Lemma \ref{L-2.3} it follows from \cite[Lemma 9.7]{Ph}
that
$$
S_M(\omega,\ffi)=\int_{\ex\cS_\gamma(\cA)}S_M(\psi,\ffi)\,d\nu_\omega(\psi)\,.
$$
This together with Lemma \ref{L-3.1} shows that
\begin{align*}
-E_{A,f}(\omega)-S_M(\omega,\ffi)
&=\int_{\ex\cS_\gamma(\cA)}(-f(\psi(A))-S_M(\psi,\ffi))\,d\nu_\omega(\psi) \\
&\le\sup\{-f(\psi(A))-S_M(\psi,\ffi):\psi\in\ex\cS_\gamma(\cA)\}\,.
\end{align*}
Therefore,
\begin{align*}
&\sup\{-E_{A,f}(\omega)-S_M(\omega,\ffi):\omega\in\cS_\gamma(\cA)\} \\
&\qquad\le\sup\{-f(\psi(A))-S_M(\psi,\ffi):\psi\in\ex\cS_\gamma(\cA)\}\,,
\end{align*}
and the converse inequality is obvious. Hence the first equality follows. The last
equality immediately follows from the definition \eqref{F-2.4}.

To prove the second equality, let $\tilde\omega$ be a maximizer of
$\omega\mapsto-f(\omega(A))-S_M(\omega,\ffi)$ on $\cS_\gamma(\cA)$. For each $m\in\bN$
with $m>\ell(A)$ we introduce a product state
$$
\psi:=\bigotimes_\bZ\tilde\omega_m
$$
of the re-localized spin chain $\bigotimes_{i\in\bZ}\cA_{[im+1,(i+1)m]}$ and define
$\bar\psi\in\cS_\gamma(\cA)$ to be the average
$$
\bar\psi:={1\over m}\sum_{k=0}^{m-1}\psi\circ\gamma^k\,.
$$
First we prove that $\bar\psi$ is $\gamma$-ergodic. For every $B_1,B_2\in\cA_{\loc}$
choose an $i_0\in\bN$ such that $B_1,B_2\in\cA_{[-i_0m,(i_0-1)m]}$. Let $n\in\bN$ be
given so that $n=jm+r$ with $j\in\bN$, $j>2i_0$ and $0\le r<m$. When $i\ge2i_0$,
$1\le t\le m$ and $0\le k\le m-1$, we have
$$
\psi(\gamma^k(B_1)\gamma^{k+im+t}(B_2))
=\psi(\gamma^k(B_1))\psi(\gamma^{k+im+t}(B_2))
=\psi(\gamma^k(B_1))\psi(\gamma^{k+t}(B_2))\,,
$$
because $\gamma^k(B_1)\in\cA_{(-\infty,i_0m]}$ and
$\gamma^{k+im+t}(B_2)\in\cA_{[(i-i_0)m+1,\infty)}$ with $i_0\le i-i_0$. Hence for
every $i\ge2i_0$ we get
\begin{align*}
\sum_{t=1}^m\bar\psi(B_1\gamma^{im+t}(B_2))
&={1\over m}\sum_{t=1}^m\sum_{k=0}^{m-1}
\psi(\gamma^k(B_1))\psi(\gamma^{k+im+t}(B_2)) \\
&=\sum_{k=0}^{m-1}\psi(\gamma^k(B_1))
\Biggl({1\over m}\sum_{t=1}^m\psi(\gamma^{k+t}(B_2))\Biggr) \\
&=\sum_{k=0}^{m-1}\psi(\gamma^k(B_1))\bar\psi(B_2)
=m\bar\psi(B_1)\bar\psi(B_2)\,.
\end{align*}
Therefore,
\begin{align*}
{1\over n}\sum_{t=1}^n\bar\psi(B_1\gamma^t(B_2))
&={1\over n}\Biggl(\sum_{t=1}^{2i_0m}+\sum_{t=jm+1}^{jm+r}\Biggr)
\bar\psi(B_1\gamma^t(B_2))
+{1\over n}\sum_{i=2i_0}^{j-1}\sum_{t=1}^m\bar\psi(B_1\gamma^{im+t}(B_2)) \\
&={1\over n}\Biggl(\sum_{t=1}^{2i_0m}+\sum_{t=jm+1}^{jm+r}\Biggr)
\bar\psi(B_1\gamma^t(B_2))
+{(j-2i_0)m\over n}\bar\psi(B_1)\bar\psi(B_2)\,,
\end{align*}
which obviously implies that
$$
\lim_{n\to\infty}{1\over n}\sum_{t=1}^n\bar\psi(B_1\gamma^t(B_2))
=\bar\psi(B_1)\bar\psi(B_2).
$$
By \cite[Theorems 4.3.17 and 4.3.22]{BR1} this is equivalent to
$\bar\psi\in\ex\cS_\gamma(\cA)$. Furthermore, since
$$
\bar\psi(A)={m-\ell(A)+1\over m}\tilde\omega(A)
+{1\over m}\sum_{k=m-\ell(A)+1}^{m-1}\psi(\gamma^k(A))\,,
$$
we get
\begin{equation}\label{F-3.3}
|\bar\psi(A)-\tilde\omega(A)|\le{2\ell(A)\|A\|\over m}\,.
\end{equation}

Now for $m$ greater than both the range of $\Phi$ and $\ell(A)$, we set a product
state
\begin{equation}\label{F-3.4}
\phi^{(m)}:=\bigotimes_\bZ\ffi_m^G
\end{equation}
of the re-localized $\bigotimes_{i\in\bZ}\cA_{[im+1,(i+1)m]}$, where $\ffi_m^G$ is the
local Gibbs state of $\cA_{[1,m]}$ for $\Phi$. We also set
\begin{align}
W&:=\sum\{\Phi(X):X\cap(-\infty,0]\ne\emptyset,\,X\cap[1,\infty)\ne\emptyset\}\,, 
\label{F-3.5}\\
K&:=\sum\{\|\Phi(X)\|:X\cap(-\infty,0]\ne\emptyset,\,X\cap[1,\infty)\ne\emptyset\}
\ (\ge\|W\|)\,. \label{F-3.6}
\end{align}
For each $j\in\bN$, since
$$
H_{jm}(\Phi)=\sum_{i=0}^{j-1}\gamma^{im}(H_m(\Phi))
+\sum_{i=1}^{j-1}\gamma^{im}(W)\,,
$$
it is clear that $\phi^{(m)}|_{\cA_{[1,jm]}}=\bigotimes_1^j\ffi_m^G$ is the perturbed
state of $\ffi_{jm}^G$ as follows:
\begin{equation}\label{F-3.7}
{\textstyle\bigotimes_1^j\ffi_m^G}=\bigl[(\ffi_{jm}^G)^{-W^{(m)}}],
\end{equation}
where $W^{(m)}:=\sum_{i=1}^{j-1}\gamma^{im}(W)$. Hence by Lemma \ref{L-2.1} and
\eqref{F-3.2} we get
\begin{align}
S(\psi_{jm},\ffi_{jm})
&\le S(\psi_{jm},\ffi_{jm}^G)+\log\lambda \nonumber\\
&\le S({\textstyle\bigotimes_1^j\tilde\omega_m},
{\textstyle\bigotimes_1^j\ffi_m^G})+2(j-1)K+\log\lambda \label{F-3.8}\\
&=jS(\tilde\omega_m,\ffi_m^G)+2(j-1)K+\log\lambda \nonumber\\
&\le jS(\tilde\omega_m,\ffi_m)+2(j-1)K+(j+1)\log\lambda\,. \nonumber
\end{align}
Since $\ffi$ can be considered as the Gibbs state for an interaction of finite range
in the re-localized $\bigotimes_{i\in\bZ}\cA_{[im+1,(i+1)m]}$, Lemma \ref{L-2.3} (the
affine property) implies that
\begin{align}
S_M(\bar\psi,\ffi)&={1\over m}\lim_{j\to\infty}{1\over j}
S(\bar\psi|_{\cA_{[1,jm]}},\ffi|_{\cA_{[1,jm]}}) \nonumber\\
&={1\over m^2}\sum_{k=0}^{m-1}\lim_{j\to\infty}{1\over j}
S(\psi\circ\gamma^k|_{\cA_{[1,jm]}},\ffi|_{\cA_{[1,jm]}}) \nonumber\\
&={1\over m}\lim_{j\to\infty}{1\over j}
S(\psi|_{\cA_{[1,jm]}},\ffi|_{\cA_{[1,jm]}}) \label{F-3.9}
\end{align}
similarly to \cite[(13.29)]{OP}. Therefore,
\begin{equation}\label{F-3.10}
S_M(\bar\psi,\ffi)\le{1\over m}S(\tilde\omega_m,\ffi_m)+{2K+\log\lambda\over m}\,.
\end{equation}
From \eqref{F-3.3} and \eqref{F-3.10} together with \eqref{F-2.1}, for any $\eps>0$
we have
$$
-f(\bar\psi(A))-S_M(\bar\psi,\ffi)
\ge-f(\tilde\omega(A))-S_M(\tilde\omega,\ffi)-\eps\,,
$$
whenever $m$ is sufficiently large. With $\bar\psi\in\ex\cS_\gamma(\cA)$ this proves
the second equality.
\endproof 

The next theorem showing the variational expression of the functional free energy
density with respect to the state $\ffi$ is a generalization of \cite[Theorem 12]{PRV}
as well as \cite[Theorem 13.11]{OP}. In fact, when $\ffi$ is a product state
$\bigotimes_\bZ\rho$ and $A$ is a one-site observable in $\cA_0$, one can easily see
that
\begin{align*}
&\max\{-f(\omega(A))-S_M(\omega,\ffi):\omega\in\cS_\gamma(\cA)\} \\
&\qquad=\max\{-f(\sigma(A))-S(\sigma,\rho):\sigma \text{ state of } \cA_0\}
\end{align*}
and
\begin{align*}
I_A(x)&=\min\{S(\sigma,\rho):\sigma \text{ state of } \cA_0,\,\sigma(A)=x\} \\
&=\sup\{-tx-\log\rho^{tA}(I):t\in\bR\}
\end{align*}
so that Theorem \ref{T-3.4}, together with Lemma \ref{L-3.3}, exactly becomes
\cite[Theorem 12]{PRV}. A typical case is the quadratic function $f(x)=x^2$, which is
familiar in quantum models of mean field type as remarked in \cite{PRV} (also in
Introduction).

The proof below is a modification of that of \cite[Theorem 13.11]{OP}. Here it should
be noted that the quantities $c(\ffi,nf(s_n(A))$ in \cite[\S13]{OP} and
$Z_\ffi(n,A,f)$ here are in the relation
$$
c(\ffi,nf(s_n(A)))=-\log Z_\ffi(n,A,f)
$$
as long as $\ffi$ is a product state.

\begin{thm}\label{T-3.4}
For every continuous $f:[\lambda_{\min}(A),\lambda_{\max}(A)] \to \bR$ the limit
$$
p_\ffi(A,f):=\lim_{n\to\infty}{1\over n}\log Z_\ffi(n,A,f)
$$
exists and
\begin{align*}
p_\ffi(A,f)
&=\sup\{-E_{A,f}(\omega)-S_M(\omega,\ffi):\omega\in\cS_\gamma(\cA)\} \\
&=\max\{-f(x)-I_A(x):x\in[\lambda_{\min}(A),\lambda_{\max}(A)]\}\,.
\end{align*}
\end{thm}

\proof 
By Lemmas \ref{L-3.2} and \ref{L-3.3} we only have to show that
$$
\limsup_{n\to\infty}{1\over n}\log Z_\ffi(n,A,f)
\le\sup\{-E_{A,f}(\omega)-S_M(\omega,\ffi):\omega\in\cS_\gamma(\cA)\}\,.
$$
To prove this, we may assume by approximation that $f$ is a polynomial. For each
$m\in\bN$ greater than both $\ell(A)$ and the range of $\Phi$, let
$\phi^{(m)}:=\bigotimes_\bZ\ffi_m^G$, a product state of the re-localized
$\bigotimes_{i\in\bZ}\cA_{[im+1,(i+1)m]}$ as in \eqref{F-3.4}. Furthermore, we set
$$
A^{(m)}:={1\over m}\sum_{k=0}^{m-\ell(A)}\gamma^k(A)\in\cA_{[1,m]}\,.
$$
According to \cite[Theorem 1]{PRV} (or \cite[Proposition 13.8]{OP}), for any $\eps>0$
there exists a symmetric (hence $\gamma^m$-invariant) state $\psi$ of
$\bigotimes_{i\in\bZ}\cA_{[im+1,(i+1)m]}$ such that
\begin{equation}\label{F-3.11}
\lim_{j\to\infty}{1\over j}\log Z_{\phi^{(m)}}^{(m)}(j,A^{(m)},mf)
<-E_{A^{(m)},mf}^{(m)}(\psi)-S_M^{(m)}(\psi,\phi^{(m)})+\eps\,,
\end{equation}
where
$$
Z_{\phi^{(m)}}^{(m)}(j,A^{(m)},mf)
:=\Tr\exp\Biggl(\log\bigl({\textstyle\bigotimes_1^jD(\ffi_m^G})\bigr)
-jmf\Biggl({1\over j}\sum_{i=0}^{j-1}\gamma^{im}(A^{(m)})\Biggr)\Biggr)\,,
$$
$$
E_{A^{(m)},mf}^{(m)}(\psi)
:=\lim_{j\to\infty}\psi\Biggl(mf\Biggl({1\over j}
\sum_{i=0}^{j-1}\gamma^{im}(A^{(m)})\Biggr)\Biggr)\,,
$$
$$
S_M^{(m)}(\psi,\phi^{(m)})
:=\lim_{j\to\infty}{1\over j}
S\bigl(\psi|_{\cA_{[1,jm]}},\phi^{(m)}|_{\cA_{[1,jm]}}\bigr)\,.
$$
Then one can define an $\omega\in\cS_\gamma(\cA)$ by
$\omega:={1\over m}\sum_{k=0}^{m-1}\psi\circ\gamma^k$. Since we assumed that $f$ is
a polynomial, there is a constant $M>0$ (depending on $\|A\|$) such that
$\|f(B_1)-f(B_2)\|\le M\|B_1-B_2\|$ for all $B_1,B_2\in\cA^{\sa}$ with
$\|B_1\|,\|B_2\|\le\|A\|$.

For each $n\in\bN$ with $n\ge m$, write $n=jm+r$ where $j\in\bN$ and $0\le r<m$. Since
$m$ is greater than the range of $\Phi$, one can write
$$
H_n(\Phi)=\sum_{i=0}^{j-1}\gamma^{im}(H_m(\Phi))
+\sum_{i=1}^{j-1}\gamma^{im}(W)+W_j\,,
$$
where $W$ is given in \eqref{F-3.5} and
$$
W_j:=\sum\{\Phi(X):X\subset[1,n],\,X\cap[jm+1,jm+r]\ne\emptyset\}\,.
$$
We have by Lemma \ref{L-2.1}
\begin{align*}
\log D(\ffi_n) &\le\log D(\ffi_n^G)+\log\lambda \\
&=-\sum_{i=0}^{j-1}\gamma^{im}(H_m(\Phi))-\sum_{i=1}^{j-1}\gamma^{im}(W)-W_j \\
&\qquad-\log\Tr\exp\Biggl(-\sum_{i=0}^{j-1}\gamma^{im}(H_m(\Phi))
-\sum_{i=1}^{j-1}\gamma^{im}(W)-W_j\Biggr)+\log\lambda \\
&\le\log\bigl({\textstyle\bigotimes_1^j D(\ffi_m^G)}\bigr)
+2jK+2\|W_j\|+\log\lambda
\end{align*}
with $K$ given in \eqref{F-3.6}. Here it is clear that $\|W_j\|\le mL$ with
$L:=\sum_{X\ni0}\|\Phi(X)\|$. Furthermore, it is readily seen that
$$
\Bigg\|s_n(A)-{1\over j}\sum_{i=0}^{j-1}\gamma^{im}(A^{(m)})\Bigg\|
\le\biggl({2\over j}+{\ell(A)\over m}\biggr)\|A\|
$$
and hence
\begin{equation}\label{F-3.12}
\Bigg\|f(s_n(A))-f\Biggl({1\over j}\sum_{i=0}^{j-1}
\gamma^{im}(A^{(m)})\Biggr)\Bigg\|
\le\biggl({2\over j}+{\ell(A)\over m}\biggr)M\|A\|\,.
\end{equation}
Therefore,
$$
\Bigg\|nf(s_n(A))-jmf\Biggl({1\over j}\sum_{i=0}^{j-1}
\gamma^{im}(A^{(m)})\Biggr)\Bigg\|
\le m\|f\|_\infty+(2m+j\ell(A))M\|A\|\,,
$$
where $\|f\|_\infty$ is the sup-norm of $f$ on
$[\lambda_{\min}(A),\lambda_{\max}(A)]$. From the above estimates we get
\begin{align*}
{1\over n}\log Z_\ffi(n,A,f)
&\le{1\over n}\log Z_{\phi^{(m)}}^{(m)}(j,A^{(m)},mf) \\
&\quad+{1\over n}\bigl\{2jK+2mL+\log\lambda
+m\|f\|_\infty+(2m+j\ell(A))M\|A\|\bigr\}
\end{align*}
so that
\begin{equation}\label{F-3.13}
\limsup_{n\to\infty}{1\over n}\log Z_\ffi(n,A,f)
\le{1\over m}\lim_{j\to\infty}{1\over j}
\log Z_{\phi^{(m)}}^{(m)}(j,A^{(m)},mf)
+{2K\over m}+{\ell(A)M\|A\|\over m}\,.
\end{equation}

Next, thanks to \eqref{F-3.12} we get
\begin{align*}
&\Bigg|\omega(f(s_n(A)))-\psi\Biggl(f\Biggl({1\over j}\sum_{i=0}^{j-1}
\gamma^{im}(A^{(m)})\Biggr)\Biggr)\Bigg| \\
&\quad\le{1\over m}\sum_{k=0}^{m-1}\|f(\gamma^k(s_n(A)))-f(s_n(A))\|
+\Bigg\|f(s_n(A))-f\Biggl({1\over j}\sum_{i=0}^{j-1}
\gamma^{im}(A^{(m)})\Biggr)\Bigg\| \\
&\quad\le{M\over m}\sum_{k=0}^{m-1}\|\gamma^k(s_n(A))-s_n(A)\|
+\biggl({2\over j}+{\ell(A)\over m}\biggr)M\|A\| \\
&\quad\le\biggl({4\over j}+{\ell(A)\over m}\biggr)M\|A\|\,.
\end{align*}
Therefore,
\begin{equation}\label{F-3.14}
\bigg|E_{A,f}(\omega)-{1\over m}E_{A^{(m)},mf}(\psi)\bigg|
\le{\ell(A)M\|A\|\over m}\,.
\end{equation}
Furthermore, we get
\begin{align*}
S(\psi_{jm},\ffi_{jm})
&\le S(\psi_{jm},\ffi_{jm}^G)+\log\lambda \\
&\le S(\psi|_{\cA_{[1,jm]}},\phi^{(m)}|_{\cA_{[1,jm]}})+2(j-1)K+\log\lambda
\end{align*}
similarly to \eqref{F-3.8} using the state perturbation \eqref{F-3.7}. Since
$$
S_M(\omega,\ffi)={1\over m}\lim_{j\to\infty}{1\over j}
S(\psi|_{\cA_{[1,jm]}},\ffi|_{\cA_{[1,jm]}})
$$
in the same way as \eqref{F-3.9}, it follows that
\begin{equation}\label{F-3.15}
S_M(\omega,\ffi)\le{1\over m}S_M^{(m)}(\psi,\phi^{(m)})+{2K\over m}\,.
\end{equation}
Inserting \eqref{F-3.13}--\eqref{F-3.15} into \eqref{F-3.11} gives
$$
\limsup_{n\to\infty}{1\over n}\log Z_\ffi(n,A,f)
\le-E_{A,f}(\omega)-S_M(\omega,\ffi)+{1\over m}(\eps+4K+2\ell(A)M\|A\|)\,,
$$
implying the required inequality because $m$ and $\eps$ are arbitrary.
\endproof 

The following is a straightforward consequence of Theorem \ref{T-3.4}.

\begin{cor}\label{C-3.5}
For every continuous $f:\bR \to \bR$, the function
$p_\ffi(\cdot\,,f)$ on $\cA_{\loc}^{\sa}$ uniquely extends to a continuous function
{\rm(}denoted by the
same $p_\ffi(\cdot\,,f)${\rm)} on $\cA^{\sa}$ satisfying
$$
p_\ffi(A,f)=\max\{-f(x)-I_A(x):x\in[\lambda_{\min}(A),\lambda_{\max}(A)]\}
$$
for all $A\in\cA^{\sa}$. Moreover, for every continuous $f,g:\bR \to \bR$ and every
$A\in\cA^{\sa}$,
$$
|p_\ffi(A,f)-p_\ffi(A,g)|
\le\max\{|f(x)-g(x)|:x\in[\lambda_{\min}(A),\lambda_{\max}(A)]\}.
$$
\end{cor}

\begin{remark}\label{R-3.6}{\rm
Suppose the ``semi-classical" case where the observable $A\in\cA_{\loc}^{\sa}$ commutes
with all $\Phi(X)$. Since $\alpha_t^\Phi(A)=A$ for all $t\in\bR$, $A$ belongs to the
centralizer of $\ffi$, i.e., $\ffi(AB)=\ffi(BA)$ for all $B\in\cA$. (To see this,
apply \cite[p.\,617]{KR} in the GNS von Neumann algebra $\pi_\ffi(\cA)''$ having the
modular automorphism group which extends $\alpha_t^\Phi$.) This implies that $s_n(A)$
commutes with $D(\ffi_n)$ for every $n\in\bN$. As stated in Remark \ref{R-2.8},
$p_\ffi(tA)$ becomes the logarithmic moment generating function of $(\mu_n)$
satisfying the large deviation principle with the good rate function $I_A(x)$ in
\eqref{F-2.4}. For any continuous $f:\bR\to\bR$ we have
$$
Z_\ffi(n,A,f)=\ffi_n(\exp(-nf(s_n(A))))
=\int_{\lambda_{\min}(A)}^{\lambda_{\max}(A)}e^{-nf(s)}\,d\mu_n(s)\,.
$$
Now Varadhan's integral lemma \cite[Theorem 4.3.1]{DZ} can be applied to obtain
$$
\lim_{n\to\infty}{1\over n}\log Z_\ffi(n,A,f)
=\max\{-f(x)-I_A(x):x\in[\lambda_{\min}(A),\lambda_{\max}(A)]\}\,.
$$
The exact large deviation principle is not formulated in our noncommutative setting
as long as the BMV-conjecture remains unsolved (see Remark \ref{R-2.8}); nevertheless
Varadhan's formula is valid as stated in Theorem \ref{T-3.4}.
}\end{remark}

\section{Perturbation of finitely correlated states}
\setcounter{equation}{0}

The notion of ($C^*$-)finitely correlated states was introduced by 
Fannes, Nachtergaele and Werner in \cite{FNW}. Let
$\cB$ be a finite-dimensional $C^*$-algebra, $\cE:\cA_0\otimes\cB\to\cB$
($\cA_0=M_d(\bC)$) a completely positive unital map and $\rho$ a state of $\cB$ such
that $\rho(\cE(I\otimes b))=\rho(b)$ for all $b\in\cB$. For each $A\in\cA_0$ define
a map $\cE_A:\cB\to\cB$ by $\cE_A(b):=\cE(A\otimes b)$, $b\in\cB$. Then the
{\it finitely correlated state} $\ffi$ determined by the triple $(\cB,\cE,\rho)$ is
the $\gamma$-invariant state of $\cA$ given by
$$
\ffi(A_0\otimes A_1\otimes\cdots\otimes A_n)
:=\rho(\cE_{A_0}\circ\cE_{A_1}\circ\cdots\circ\cE_{A_n}(\1_\cB))
\qquad (A_i\in\cA_i\,,\ 0\le i\le n)\,.
$$
As was shown in the proof of \cite[Proposition 4.4]{HMO}, a finitely correlated state
has the following upper factorization property, which will be useful in our
discussions below.

\begin{lemma}\label{L-4.1}
If $\ffi$ is a finitely correlated state of $\cA$, then there exists a constant
$\alpha\ge1$ such that
$$
\ffi\le\alpha\bigl(\ffi|_{\cA_{(-\infty,0]}}\bigr)
\otimes\bigl(\ffi|_{\cA_{[1,\infty)}}\bigr)\,.
$$
\end{lemma}

The next proposition is a generalization of \cite[Proposition 11.2]{OP}.

\begin{prop}\label{P-4.2}
Let $\ffi$ be a finitely correlated state of $\cA$. For every
$\omega\in\cS_\gamma(\cA)$ the mean relative entropy
$$
S_M(\omega,\ffi)=\lim_{n\to\infty}{1\over n}S(\omega_n,\ffi_n)
$$
exists. Moreover, the function $\omega\in\cS_\gamma(\cA)\mapsto S_M(\omega,\ffi)$ is
affine and weakly* lower semicontinuous on $\cS_\gamma(\cA)$.
\end{prop}

\proof 
The proof of the first assertion is a slight modification of that of
\cite[Theorem 2.1]{HP2} while it will be repeated below for the convenience of the
remaining proof. For each $n,m\in\bN$ with $n\ge m$, write $n=jm+r$ with $j\in\bN$
and $0\le r<m$. Lemma \ref{L-4.1} implies that
\begin{equation}\label{F-4.1}
\ffi_n\le\alpha^j\Biggl(\bigotimes_{i=0}^{j-1}
\bigl(\ffi|_{\cA_{[im+1,(i+1)m]}}\bigr)\Biggr)
\otimes\bigl(\ffi|_{\cA_{[jm+1,jm+r]}}\bigr)\,.
\end{equation}
Consider the product state $\phi^{(m)}:=\bigotimes_\bZ\ffi_m$ of the re-localized spin
chain $\bigotimes_{i\in\bZ}\cA_{[im+1,(i+1)m]}$. For every $\omega\in\cS_\gamma(\cA)$
we have
\begin{equation}\label{F-4.2}
S(\omega_n,\ffi_n)\ge S(\omega_{jm},\ffi_{jm})
\ge S(\omega_{jm},{\textstyle\bigotimes_1^j\ffi_m})-j\log\alpha
\end{equation}
due to the monotonicity of relative entropy and \eqref{F-4.1}. Dividing \eqref{F-4.2}
by $n$ and letting $n\to\infty$ with $m$ fixed we get
$$
\liminf_{n\to\infty}{1\over n}S(\omega_n,\ffi_n)
\ge{1\over m}S_M^{(m)}(\omega,\phi^{(m)})-{\log\alpha\over m}\,,
$$
where $S_M^{(m)}(\omega,\phi^{(m)})$ denotes the mean relative entropy in the
re-localized $\bigotimes_{i\in\bZ}\cA_{[im+1,(i+1)m]}$ as in \eqref{F-3.11}. Since
$S_M^{(m)}(\omega,\phi^{(m)})\ge S(\omega_m,\ffi_m)$ by \cite[(2.1)]{HP0}, we
further get
$$
\liminf_{n\to\infty}{1\over n}S(\omega_n,\ffi_n)
\ge{1\over m}S(\omega_m,\ffi_m)-{\log\alpha\over m}\,.
$$
Since $m\in\bN$ is arbitrary, this shows the existence of $S_M(\omega,\ffi)$ and the
above inequalities become
\begin{equation}\label{F-4.3}
S_M(\omega,\ffi)\ge{1\over m}S(\omega_m,\ffi_m)-{\log\alpha\over m}\,.
\end{equation}

The affinity of $\omega\mapsto S_M(\omega,\ffi)$ is a consequence of the general
property \cite[Proposition 5.24]{OP}. Assume that
$\omega,\omega^{(k)}\in\cS_\gamma(\cA)$ and $\omega^{(k)}\to\omega$ weakly*. Then
from \eqref{F-4.3} we have
$$
\liminf_{k\to\infty}S_M(\omega^{(k)},\ffi)
\ge{1\over m}\liminf_{k\to\infty}S(\omega_m^{(k)},\ffi_m)
-{\log\alpha\over m}
\ge{1\over m}S(\omega_m,\ffi_m)-{\log\alpha\over m}
$$
thanks to the lower semicontinuity of relative entropy (in fact,
$\omega\mapsto S(\omega_m,\ffi_m)$ is continuous due to finite dimensionality).
Letting $m\to\infty$ shows the lower semicontinuity of
$\omega\mapsto S_M(\omega,\ffi)$. 
\endproof 

Next we show the existence of the free energy density with respect to a finitely
correlated state $\ffi$. Since $\ffi$ is not assumed to be locally faithful in the
sense that $D(\ffi_n)$ is strictly positive for every $n\in\bN$, we need to be
careful in defining $\Tr\exp\bigl(\log D(\ffi_n)-B\bigr)$ for $B\in\cA_{[1,n]}^{\sa}$.
Let $D$ be a nonzero positive semidefinite matrix and $B$ a Hermitian matrix in
$M_N(\bC)$. It is known \cite[Lemma 4.1]{HP1} that
$$
\lim_{\eps\searrow0}e^{\log(D+\eps I)-B}=P(e^{P(\log D)P-PBP})P\,,
$$
where $P$ is the support projection of $D$. Hence one can define $\Tr\,e^{\log D-B}$
by
\begin{equation}\label{F-4.4}
\Tr\,e^{\log D-B}:=\lim_{\eps\searrow0}\Tr e^{\log(D+\eps I)-B}
=\Tr Pe^{P(\log D)P-PBP}\,.
\end{equation}

\begin{prop}\label{P-4.3}
Let $\ffi$ be a finitely correlated state of $\cA$. For every $A\in\cA_{\loc}^{\sa}$ so
that $A\in\cA_\Lambda^{\sa}$ with a finite $\Lambda\subset\bZ$, the free energy density
$$
p_\ffi(A):=\lim_{n\to\infty}{1\over n}
\log\Tr\exp\Biggl(\log D(\ffi_n)
-\sum_{\Lambda+k\subset[1,n]}\gamma^k(A)\Biggr)
$$
exists {\rm(}independently of the choice of $\Lambda${\rm)}. Moreover, $p_\ffi$ is
convex and Lipschitz continuous with $|p_\ffi(A)-p_\ffi(B)|\le \|A-B\|$, and
therefore it uniquely extends to a function on $\cA^{\sa}$ with the same properties.
\end{prop}

\proof
To prove the first assertion we may assume that $A\in\cA_{[1,\ell(A)]}^{\sa}$ with
some $\ell(A)\in\bN$. For each $n,m\in\bN$ with $n\ge m>\ell(A)$, write $n=jm+r$ with
$0\le r<m$. From \eqref{F-4.1} we get
$$
D(\ffi_n)\le\alpha^j\prod_{i=0}^{j-1}\gamma^{im}(D(\ffi_m))
$$
with a constant $\alpha\ge1$ independent of $n,m$. For any $\eps>0$ this implies that
$$
D(\ffi_n)+\eps^jI\le\alpha^j\prod_{i=0}^{j-1}\gamma^{im}(D(\ffi_m)+\eps I)\,.
$$
Furthermore, it is immediately seen that
$$
\sum_{k=0}^{n-\ell(A)}\gamma^k(A)\ge\sum_{i=0}^{j-1}
\gamma^{im}\Biggl(\sum_{k=0}^{m-\ell(A)}\gamma^k(A)\Biggr)
-(j(\ell(A)-1)+r)\|A\|\,.
$$
Set $h_n:=\sum_{k=0}^{n-\ell(A)}\gamma^k(A)$. From the above two inequalities we get
\begin{align*}
&\Tr\exp\bigl(\log(D(\ffi_n)+\eps^jI)-h_n\bigr) \\
&\quad\le\bigl\{\Tr\exp\bigl(\log(D(\ffi_m)+\eps I)-h_m\bigr)\bigr\}^j
\exp\bigl(j\log\alpha+(j(\ell(A)-1)+r)\|A\|\bigr).
\end{align*}
In view of the definition \eqref{F-4.4}, letting $\eps\searrow0$ gives
\begin{align*}
&\Tr\exp\bigl(\log D(\ffi_n)-h_n\bigr) \\
&\quad\le\bigl\{\Tr\exp\bigl(\log D(\ffi_m)-h_m\bigr)\bigr\}^j
\exp\bigl(j\log\alpha+(j(\ell(A)-1)+r)\|A\|\bigr)
\end{align*}
so that
\begin{align*}
&\limsup_{n\to\infty}{1\over n}
\log\Tr\exp\bigl(\log D(\ffi_n)-h_n\bigr) \\
&\quad\le{1\over m}\log\Tr\exp\bigl(\log D(\ffi_m)-h_m\bigr)
+{\log\alpha\over m}+{\ell(A)-1\over m}\|A\|\,.
\end{align*}
Since $m$ ($>\ell(A)$) is arbitrary, this shows the existence of the limit
$p_\ffi(A)$. It is obvious that $p_\ffi(A)$ is independent of the choice of $\Lambda$.
It is also clear that $p_\ffi(A)$ on $\cA_{\loc}^{\sa}$ is convex and satisfies
$|p_\ffi(A)-p_\ffi(B)|\le\|A-B\|$ for all $A,B\in\cA_{\loc}^{\sa}$, from which the
second part of the proposition follows.
\endproof 

\begin{remark}\label{R-4.4}{\rm
The limit $\tilde p_\ffi(A)$ similar to $p_\ffi(A)$ was referred to in Remark
\ref{R-2.7} from the viewpoint of large deviations. In \cite{HMO} the limit
$\tilde p_\ffi(A)$ was shown to exist for any $A\in\cA_{\loc}^{\sa}$ when $\ffi$ is a
finitely correlated state (as well as when $\ffi$ is a Gibbs state). The proof for
$\tilde p_\ffi(A)$ is more involved than the above for $p_\ffi(A)$ and relies on the
estimate in \cite[Theorem 3.7]{LR} related to Gibbs state perturbation.
}\end{remark}

Once we had Propositions \ref{P-4.2} and \ref{P-4.3}, it is natural to expect that
$S_M(\omega,\ffi)$ and $p_\ffi(A)$ enjoy the same Legendre transform formulas as
(4) and (5) of Corollary \ref{C-2.5} in the Gibbs state case. But this is still
unsolved while the following inequality is easy as Lemma \ref{L-3.2}. For the proof
use \cite[(4.2)]{HP2} or \cite[Proposition 1.11]{OP}, the extended version of
\eqref{F-3.1}.

\begin{prop}\label{P-4.5}
Let $\ffi$ be a finitely correlated state of $\cA$. For every $A\in\cA^{\sa}$,
$$
p_\ffi(A)\ge\max\{-\omega(A)-S_M(\omega,\ffi):\omega\in\cS_\gamma(\cA)\}\,.
$$
\end{prop}

\begin{remark}\label{R-4.5}{\rm
Suppose that $\ffi$ satisfies the lower factorization property
$$
\ffi\ge\beta\bigl(\ffi|_{\cA_{(-\infty,0]}}\bigr)
\otimes\bigl(\ffi|_{\cA_{[1,\infty)}}\bigr)
$$
for some $\beta>0$ (the opposite version of Lemma \ref{L-4.1}). (In fact, it is
enough to suppose a slightly weaker version of lower factorization as in
\cite[Definition 4.1]{HMO}.) Then all the results in Section 3 are true for $\ffi$.
The proofs can be carried out  similarly to those in Section 3; in fact, they are
even easier without the state perturbation technique. However, the lower factorization
property for finitely correlated states is quite strong; for example, one can easily
see that a classical irreducible Markov chain has this property if and only if its
transition stochastic matrix is strictly positive (i.e., all entries are strictly
positive), which is stronger than the strong mixing property. More details are in
\cite{HMO}.
}\end{remark}

In the rest of this section, we assume that $\ffi$ is a $\gamma$-invariant
{\it quantum Markov state} of Accardi and Frigerio type \cite{AF}, and further assume
that $\ffi$ is locally faithful. According to \cite{AL,Oh}, there exists a conditional
expectation $E$ from $M_d(\bC)\otimes M_d(\bC)$ into $M_d(\bC)$ such that
$\ffi_0\circ E(I\otimes A)=\ffi_0(A)$ for all $A\in M_d(\bC)$ and
$$
\ffi(A_0\otimes A_1\otimes\dots\otimes A_n)
=\ffi_0(E(A_0\otimes E(A_1\otimes\dots\otimes E(A_{n-1}\otimes A_n)\cdots)))
$$
for all $A_0,A_1,\dots,A_n\in M_d(\bC)$, where $\ffi_0:=\ffi|_{\cA_0}$. Set
$\cB:=E(M_d(\bC)\otimes M_d(\bC))$, a subalgebra of $M_d(\bC)$,
$\cE:=E|_{M_d(\bC)\otimes\cB}$ and $\rho:=\ffi_0|_\cB$. Then $\ffi$ is a
finitely correlated state with the triple $(\cB,\cE,\rho)$. Let $q_1,\dots,q_k$ be
the minimal central projections of $\cB$; then $\cB q_i\cong M_{d_i}(\bC)$ and $\cB$
is decomposed as
$$
\cB=\bigoplus_{i=1}^k\cB q_i
=\bigoplus_{i=1}^k\bigl(M_{d_i}(\bC)\otimes I_{m_i}\bigr)\,,
$$
where $m_i$ is the multiplicity of $M_{d_i}(\bC)$ in $M_d(\bC)$. Let $\cB'$ be the
relative commutant of $\cB$ in $M_d(\bC)$ so that
$\cB'=\bigoplus_{i=1}^kI_{d_i}\otimes M_{m_i}(\bC)$. For each $m,n\in\bZ$, $m\le n$,
set
$$
\widetilde\cA_{[m,n]}:=\cB'\otimes\cA_{[m+1,n-1]}\otimes\cB
\ \ (\subset\cA_{[m,n]})
$$
with convention $\widetilde\cA_{[n,n]}:=\bC I$ ($\subset\cA_n$). Let
$\mathcal{C}:=\bigoplus_{i=1}^kM_{d_i}(\bC)\otimes M_{m_i}(\bC)$ ($\subset M_d(\bC)$)
and $E_{\mathcal{C}}$ be the pinching
$A\in M_d(\bC)\mapsto\sum_{i=1}^kq_iAq_i\in\mathcal{C}$ (or the conditional
expectation onto $\mathcal{C}$ with respect to the trace). The following properties
were shown in \cite{AL,Oh}:
\begin{itemize}
\item[\rm(i)] There exist positive linear functionals $\rho_{ij}$ on
$M_{m_i}(\bC)\otimes M_{d_j}(\bC)$, $1\le i,j\le k$, such that
$$
\cE=\Biggl(\bigoplus_{i,j=1}^k\mathrm{id}_{M_{d_i}(\bC)}
\otimes\rho_{ij}\Biggr)\circ
(E_{\mathcal{C}}\otimes\mathrm{id}_\cB)\,.
$$
\item[\rm(ii)] Let $T_{ij}$ be the density matrices of $\rho_{ij}$ for $1\le i,j\le k$.
Then the density matrix of $\ffi|_{\widetilde\cA_{[m,n]}}$ is
\begin{equation}\label{F-4.5}
\widetilde D_{[m,n]}:=\bigoplus_{i_m,i_{m+1},\dots,i_n}\rho(q_{i_m})
T_{i_mi_{m+1}}\otimes T_{i_{m+1}i_{m+2}}\otimes\dots\otimes T_{i_{n-1}i_n}\,.
\end{equation}
\end{itemize}

The density matrices $\widetilde D_{[m,n]}$ have a simple form of product type. Since
$T_{ij}$ is strictly positive in $M_{m_i}(\bC)\otimes M_{d_j}(\bC)$ for each $i,j$
due to the local faithfulness of $\ffi$, a $\gamma$-invariant nearest-neighbor
interaction $\Phi$ can be defined by
$$
\Phi([0,1]):=-\sum_{i,j=1}^k\log T_{ij}\in\cB'\otimes\cB\subset\cA_{[0,1]}\,,
\quad\Phi([n,n+1]):=\gamma^n(\Phi([0,1]))\,.
$$
Then the density of the local Gibbs state of $\cA_{[m,n]}$ for $\Phi$ is
$$
\bigoplus_{i_m,\dots,i_n}T_{i_mi_{m+1}}\otimes\dots\otimes T_{i_{n-1}i_n}\,,
$$
and the automorphism group $\alpha_t^\Phi$ is given by
\begin{equation}\label{F-4.6}
\alpha_t^\Phi(A)=\lim_{m\to-\infty,n\to\infty}
\widetilde D_{[m,n]}^{-it}A\widetilde D_{[m,n]}^{it}
\qquad (A\in\cA)\,.
\end{equation}
Hence $\ffi$ is the $\alpha^\Phi$-KMS state (or the Gibbs state for $\Phi$) and so all
the results in Sections 2 and 3 hold for $\ffi$. Below let us further investigate the
relation between $p_\ffi(A)$ in \eqref{F-2.3} and $\tilde p_\ffi(A)$ in \eqref{F-2.5}.

The centralizer of $\ffi$ is given by
$$
\cA_\ffi:=\{A\in\cA:\ffi(AB)=\ffi(BA)\ \mbox{for all $B\in\cA$}\}\,,
$$
which is a $\gamma$-invariant $C^*$-subalgebra of $\cA$. For each $m,n\in\bZ$ with
$m\le n$, we also define
$$
(\widetilde\cA_{[m,n]})_\ffi:=\{A\in\widetilde\cA_{[m,n]}:\ffi(AB)=\ffi(BA)
\ \mbox{for all $B\in\widetilde\cA_{[m,n]}$}\}\,.
$$

\begin{lemma}\label{L-4.7}
If $m'\le m\le n\le n'$ in $\bZ$, then
$(\widetilde\cA_{[m,n]})_\ffi\subset(\widetilde\cA_{[m',n']})_\ffi\subset\cA_\ffi$.
Moreover,
$\widetilde\cA_{\ffi,\loc}:=\bigcup_{n=1}^\infty(\widetilde\cA_{[-n,n]})_\ffi$ is a
dense $*$-subalgebra of $\cA_\ffi$.
\end{lemma}

\proof
Since $(\widetilde\cA_{[m,n]})_\ffi$ is the relative commutant of
$\{\widetilde D_{[m,n]}\}$ in $\widetilde\cA_{[m,n]}$, the first assertion is
immediately seen from the form \eqref{F-4.5} of $\widetilde D_{[m,n]}$. Furthermore,
it is also obvious from \eqref{F-4.6} that
$\alpha_t^\Phi(\widetilde\cA_{[m,n]})=\widetilde\cA_{[m,n]}$, $t\in\bR$, for any
$m\le n$. By \cite{Ta} applied in the GNS von Neumann algebra $\pi_\ffi(\cA)''$ with
the modular automorphism group extending $\alpha_t^\Phi$, there exists the conditional
expectation $E_{[m,n]}:\cA\to\widetilde\cA_{[m,n]}$ with $\ffi\circ E_{[m,n]}=\ffi$.
Then it is clear that $\|E_{[m,n]}(A)-A\|\to0$ as $m\to-\infty$ and $n\to\infty$ for
any $A\in\cA$. Now let $A\in\cA_\ffi$. Since
$$
\ffi(E_{[m,n]}(A)B)=\ffi(AB)=\ffi(BA)=\ffi(BE_{[m,n]}(A))\,,
\qquad B\in\widetilde\cA_{[m,n]}\,,
$$
we have $E_{[m,n]}(A)\in(\widetilde\cA_{[m,n]})_\ffi$ for any $m\le n$, implying the
latter assertion.
\endproof 

\begin{lemma}\label{L-4.8}
For every $\omega\in\cS_\gamma(\cA)$,
$$
S_M(\omega,\ffi)=\lim_{n\to\infty}{1\over n}
S\bigl(\omega|_{(\widetilde\cA_{[1,n]})_\ffi},
\ffi|_{(\widetilde\cA_{[1,n]})_\ffi}\bigr)
$$
and hence $S_M(\omega,\ffi)$ is determined by $\omega|_{\cA_\ffi}$. Moreover, if
$\omega,\omega^{(i)}\in\cS_\gamma(\cA)$, $i\in\bN$, and
$\omega^{(i)}|_{\cA_\ffi}\to\omega|_{\cA_\ffi}$ in the weak* topology, then
$$
S_M(\omega,\ffi)\le\liminf_{i\to\infty}S_M(\omega^{(i)},\ffi)\,.
$$
\end{lemma}

\proof 
The proof of the first assertion is essentially the same as that of
\cite[Theorem 2.1]{HP0} as will be sketched below. Let
$T_{ij}=\sum_{\ell=1}^{L_{ij}}\lambda_{ij\ell}e_{ij\ell}$ be the spectral
decomposition of $T_{ij}$ for $1\le i,j\le k$, and $\Theta$ be the set of all
$(i,j,\ell)$ with $1\le i,j\le k$ and $1\le\ell\le L_{ij}$. For each $n\in\bN$ let
$K_n$ be the set of all tuples $(n_\theta)_{\theta\in\Theta}$ of nonnegative integers
such that $\sum_{\theta\in\Theta}n_\theta=n-1$. For each $1\le i\le k$ and
$(n_\theta)\in K_n$ we denote by $I_{i,(n_\theta)}$ the set of all
$(i_1,i_2,\dots,i_n;\ell_1,\ell_2,\dots,\ell_{n-1})$ such that $i_1=i$ and
$\#\{r\in[1,n-1]:(i_r,i_{r+1},\ell_r)=\theta\}=n_\theta$ for all $\theta\in\Theta$,
and define the projection $P_{i,(n_\theta)}$ in $\widetilde\cA_{[1,n]}$ and
$\lambda_{i,(n_\theta)}\in\bR$ by
\begin{align*}
P_{i,(n_\theta)}
&:=\sum_{(i_1,\dots,i_n;\ell_1,\dots,\ell_{n-1})\in I_{i,(n_\theta)}}
e_{i_1i_2\ell_1}\otimes e_{i_2i_3\ell_2}\otimes\dots\otimes
e_{i_{n-1}i_n\ell_{n-1}}\,, \\
\lambda_{i,(n_\theta)}&:=\rho(q_i)\prod_{\theta\in\Theta}\lambda_\theta^{n_\theta}
\quad\mbox{where\quad$\lambda_\theta:=\lambda_{ij\ell}$
\ \ for\ \ $\theta=(i,j,\ell)$}\,.
\end{align*}
Then $\sum_{i=1}^k\sum_{(n_\theta)\in K_n}P_{i,(n_\theta)}=I$ and
$\widetilde D_{[1,n]}$ is written as
$$
\widetilde D_{[1,n]}=\sum_{i=1}^k\sum_{(n_\theta)\in K_n}
\lambda_{i,(n_\theta)}P_{i,(n_\theta)}\,.
$$
Now, for each $\omega\in\cS_\gamma(\cA)$, the proof of \cite[Theorem 2.1]{HP0} implies
that
$$
S(\omega_{n-2},\ffi_{n-2})
\le S\bigl(\omega|_{\widetilde\cA_{[1,n]}},
\ffi|_{\widetilde\cA_{[1,n]}}\bigr)
\le S\bigl(\omega|_{(\widetilde\cA_{[1,n]})_\ffi},
\ffi|_{(\widetilde\cA_{[1,n]})_\ffi}\bigr)+\log k+\log\#K_n
$$
for every $n\ge3$. Since $\#K_n\le n^{\#\Theta}$, we get
$$
S_M(\omega,\ffi)\le\liminf_{n\to\infty}{1\over n}
S\bigl(\omega|_{(\widetilde\cA_{[1,n]})_\ffi},
\ffi|_{(\widetilde\cA_{[1,n]})_\ffi}\bigr)\,,
$$
which proves the first assertion.

Set $\gamma:=1/\min_{1\le i\le k}\rho(q_i)$. For each $m,m'\in\bN$, since it follows
from \eqref{F-4.5} that
$$
\ffi|_{(\widetilde\cA_{[1,m]})_\ffi\otimes(\widetilde\cA_{[m+1,m+m']})_\ffi}
\le\gamma\bigl(\ffi|_{(\widetilde\cA_{[1,m]})_\ffi}\bigr)
\otimes\bigl(\ffi|_{(\widetilde\cA_{[m+1,m+m']})_\ffi}\bigr)\,,
$$
we get
\begin{align*}
&S\bigl(\omega|_{(\widetilde\cA_{[1,m+m']})_\ffi},
\ffi|_{(\widetilde\cA_{[1,m+m']})_\ffi}\bigr) \\
&\quad\ge S\bigl(\omega|_{(\widetilde\cA_{[1,m]})_\ffi\otimes
(\widetilde\cA_{[m+1,m+m']})_\ffi},
\ffi|_{(\widetilde\cA_{[1,m]})_\ffi\otimes
(\widetilde\cA_{[m+1,m+m']})_\ffi}\bigr)
-\log\gamma \\
&\quad\ge S\bigl(\omega|_{(\widetilde\cA_{[1,m]})_\ffi},
\ffi|_{(\widetilde\cA_{[1,m]})_\ffi}\bigr)
+S\bigl(\omega|_{(\widetilde\cA_{[1,m']})_\ffi},
\ffi|_{(\widetilde\cA_{[1,m']})_\ffi}\bigr)-\log\gamma
\end{align*}
due to the monotonicity and the superadditivity of relative entropy
\cite[Corollary 5.21]{OP}. Let $\omega$ and $\omega^{(i)}$ be given as stated in the
lemma. For any $m\in\bN$ and $n=jm+r$ with $j\in\bN$ and $0\le r<m$, the above
inequality gives
$$
S\bigl(\omega^{(i)}|_{(\widetilde\cA_{[1,n]})_\ffi},
\ffi|_{(\widetilde\cA_{[1,n]})_\ffi}\bigr) \\
\ge jS\bigl(\omega^{(i)}|_{(\widetilde\cA_{[1,m]})_\ffi},
\ffi|_{(\widetilde\cA_{[1,m]})_\ffi}\bigr)-j\log\gamma\,.
$$
Dividing this by $n$ and letting $n\to\infty$ with $m$ fixed we get
$$
S_M(\omega^{(i)},\ffi)
\ge{1\over m}S\bigl(\omega^{(i)}|_{(\widetilde\cA_{[1,m]})_\ffi},
\ffi|_{(\widetilde\cA_{[1,m]})_\ffi}\bigr)-{\log\gamma\over m}
$$
and hence
$$
\liminf_{i\to\infty}S_M(\omega^{(i)},\ffi)
\ge{1\over m}S\bigl(\omega|_{(\widetilde\cA_{[1,m]})_\ffi},
\ffi|_{(\widetilde\cA_{[1,m]})_\ffi}\bigr)-{\log\gamma\over m}\,.
$$
Letting $m\to\infty$ gives the latter assertion.
\endproof 

In addition to the variational expression in Corollary \ref{C-2.5}\,(5) we have

\begin{thm}\label{T-4.9}
For every $\omega\in\cS_\gamma(\cA)$,
\begin{align*}
S_M(\omega,\ffi)&=\sup\{-\omega(A)-p_\ffi(A):A\in\cA_\ffi^{\sa}\} \\
&=\sup\{-\omega(A)-\tilde p_\ffi(A):A\in\cA_{\loc}^{\sa}\}\,,
\end{align*}
where $\tilde p_\ffi(A)$ is given in \eqref{F-2.5}.
\end{thm}

\proof 
The proof of the first equality is a simple duality argument. Set
$\Gamma:=\{\omega|_{\cA_\ffi^{\sa}}:\omega\in\cS_\gamma(\cA)\}$, which is a weakly*
compact and convex subset of $(\cA_\ffi^{\sa})^*$, the dual Banach space of the real
Banach space $\cA_\ffi^{\sa}$. From Lemma \ref{L-4.8} one can define
$F:(\cA_\ffi^{\sa})^*\to[0,+\infty]$ by
$$
F(f):=\begin{cases}
S_M(\omega,\ffi)
& \text{if } f=\omega|_{\cA_\ffi^{\sa}} \text{ with some }\omega\in\cS_\gamma(\cA), \\
+\infty & \text{otherwise},
\end{cases}
$$
which is affine and weakly* lower semicontinuous on $(\cA_\ffi^{\sa})^*$ by Proposition
\ref{P-4.2} and Lemma \ref{L-4.8}. Corollary \ref{C-2.5}\,(4) says that
$$
p_\ffi(A)=\max\{-f(A)-F(f):f\in(\cA_\ffi^{\sa})^*\}\,,
\qquad A\in\cA_\ffi^{\sa}\,.
$$
Hence it follows by duality \cite[Proposition I.4.1]{ET} that
$$
F(f)=\sup\{-f(A)-p_\ffi(A):A\in\cA_\ffi^{\sa}\}\,,
\qquad f\in(\cA_\ffi^{\sa})^*\,.
$$
For every $\omega\in\cS_\gamma(\cA)$ this means the first equality, which also gives
\begin{equation}\label{F-4.7}
S_M(\omega,\ffi)=\sup\{-\omega(A)-p_\ffi(A):A\in\widetilde\cA_{\ffi,\loc}^{\sa}\}
\end{equation}
thanks to Lemma \ref{L-4.7}.

To prove the second equality, we show that $p_\ffi(A)=\tilde p_\ffi(A)$ for all
$A\in\widetilde\cA_{\ffi,\loc}^{\sa}$. Thanks to Lemma \ref{L-4.7} and the
$\gamma$-invariance of $p_\ffi$ and $\tilde p_\ffi$,  we may assume that
$A\in(\widetilde\cA_{[1,m]})_\ffi^\sa$ for some $m\in\bN$. For each $n\in\bN$ and
$0\le k\le n-m$, we have
$\gamma^k(A)\in(\widetilde\cA_{[1+k,m+k]})_\ffi\subset(\widetilde\cA_{[1,n]})_\ffi$
so that $\exp\bigl(-\sum_{k=0}^{n-m}\gamma^k(A)\bigr)\in(\widetilde\cA_{[1,n]})_\ffi$.
Furthermore, since
$\widetilde\cA_{[1,n]}\subset\cA_{[1,n]}\subset\widetilde\cA_{[0,n+1]}$, it is easy to
see by Lemma 4.7 that $(\widetilde\cA_{[1,n]})_\ffi\subset(\cA_{[1,n]})_\ffi$. Hence
we get $\exp\bigl(-\sum_{k=0}^{n-m}\gamma^k(A)\bigr)\in(\cA_{[1,n]})_\ffi$, which
implies that $\exp\bigl(-\sum_{k=0}^{n-m}\gamma^k(A)\bigr)$ commutes with the density
$D(\ffi_n)$ so that
$$
\ffi\Biggl(\exp\Biggl(-\sum_{k=0}^{n-m}\gamma^k(A)\Biggr)\Biggr)
=\Tr\exp\Biggl(\log D(\ffi_n)-\sum_{k=0}^{n-m}\gamma^k(A)\Biggr)\,,
$$
showing $p_\ffi(A)=\tilde p_\ffi(A)$ by definitions \eqref{F-2.3} and
\eqref{F-2.5}.
From this and \eqref{F-4.7} we get
\begin{align*}
S_M(\omega,\ffi)&\le\sup\{-\omega(A)-\tilde p_\ffi(A):A\in\cA_{\loc}^{\sa}\} \\
&\le\sup\{-\omega(A)-p_\ffi(A):A\in\cA_{\loc}^{\sa}\}=S_M(\omega,\ffi)
\end{align*}
thanks to \eqref{F-2.6} and Corollary \ref{C-2.5}\,(5), implying the second equality.
\endproof 

\begin{cor}\label{C-4.10}
The function $p_\ffi$ on $\cA^{\sa}$ is the lower semicontinuous convex envelope of
$\tilde p_\ffi$ on $\cA_{\loc}^{\sa}$ in the sense that $p_\ffi$ is the largest among
lower semicontinuous and convex functions $q$ on $\cA^{\sa}$ satisfying
$q\le\tilde p_\ffi$ on $\cA_{\loc}^{\sa}$.
\end{cor}

\proof 
Let $q$ be as stated in the corollary. Define $Q:(\cA^{\sa})^*\to(-\infty,+\infty]$ by
$$
Q(f):=\sup\{-f(A)-q(A):A\in\cA^{\sa}\}
\qquad (f\in(\cA^{\sa})^*)\,.
$$
Let us prove that
\begin{equation}\label{F-4.8}
\begin{cases}
Q(\omega)\ge S_M(\omega,\ffi) & \text{if $\omega\in\cS_\gamma(\cA)$}, \\
Q(f)=+\infty & \text{if $f\in(\cA^{\sa})^*\setminus\cS_\gamma(\cA)$}.
\end{cases}
\end{equation}
For $\omega\in\cS_\gamma(\cA)$ Theorem \ref{T-4.9} gives
$$
Q(\omega)\ge\sup\{-\omega(A)-\tilde p_\ffi(A):A\in\cA_{\loc}^{\sa}\}
=S_M(\omega,\ffi)\,.
$$
For $f\in(\cA^{\sa})^*\setminus\cS_\gamma(\cA)$ we may consider
the following three cases:
\begin{itemize}
\item[\rm(a)] $f(A)<0$ for some positive $A\in\cA_{\loc}$,
\item[\rm(b)] $f(\1)\ne1$,
\item[\rm(c)] $f(A)\ne f(\gamma(A))$ for some $A\in\cA^{\sa}$.
\end{itemize}
In case (a), since $q(\alpha A)\le\tilde p_\ffi(\alpha A)\le0$ for $\alpha>0$, we
get $-f(\alpha A)-q(\alpha A)\ge-\alpha f(A)\to+\infty$ as $\alpha\to+\infty$. In
case (b), since $q(\alpha\1)\le\tilde p_\ffi(\alpha\1)=-\alpha$, we get
$-f(\alpha\1)-q(\alpha\1)\ge-\alpha(f(\1)-1)\to+\infty$ as $\alpha\to+\infty$ or
$-\infty$ according as $f(\1)<1$ or $f(\1)>1$. Finally in case (c), since
$$
q(\alpha(A-\gamma(A)))\le\tilde p_\ffi(\alpha(A-\gamma(A)))
=\lim_{n\to\infty}{1\over n}\log\ffi(e^{-\alpha(A-\gamma^n(A))})=0\,,
$$
we get
$-f(\alpha(A-\gamma(A)))-q(\alpha(A-\gamma(A)))\ge-\alpha f(A-\gamma(A))\to+\infty$ as
$\alpha\to+\infty$ or $-\infty$ according as $f(A)<f(\gamma(A))$ or
$f(A)>f(\gamma(A))$. Hence \eqref{F-4.8} follows. By duality this implies that
$q\le p_\ffi$ on $\cA^{\sa}$.
\endproof 

In particular, when $\ffi$ is the product state $\bigotimes_\bZ\rho$ of a not
necessarily faithful $\rho$, all the variational expressions in Corollary \ref{C-2.5}
and Theorem \ref{T-4.9} are valid for $\ffi$, and so Corollary \ref{C-4.10} holds for
$\ffi$. Although we have no strong evidence, it might be conjectured that Corollary
\ref{C-4.10} is true generally for the Gibbs-KMS state $\ffi$ treated in Sections 2
and 3.

\section{Concluding remarks: guide to the case of arbitrary dimension}
\setcounter{equation}{0}

In this paper we confined ourselves to the one-dimensional spin chain case for the
following reasons. First, our main motivation came from recent developments on large
deviations in spin chains, where the differentiability of logarithmic moment generating
functions is crucial. The corresponding functions in our setting are free energy
density functions so that we wanted to provide their differentiability (see Theorem
\ref{T-2.4}\,(c) and Corollary \ref{C-2.5}\,(2)), and the one-dimensionality is
essential for this. Secondly, finitely correlated states treated in the latter half
are defined only in a one-dimensional spin chain though some attempts to
multi-dimensional extension were made for similar states of quantum Markov type (see
\cite{AcFi1,AcFi2} for example). However, all the discussions (except the
differentiability assertions) presented for a Gibbs state of one-dimension in Sections
2 and 3 can also work well in the setting of arbitrary dimension but in high
temperature regime, which we outline below.

Consider a $\nu$-dimensional spin chain $\cA:=\bigotimes_{k\in\bZ^\nu}\cA_k$,
$\cA_k=M_d(\bC)$, with the translation automorphism group $\gamma_k$, $k\in\bZ^\nu$,
and local algebras $\cA_\Lambda:=\bigotimes_{k\in\Lambda}\cA_k$ for finite
$\Lambda\subset\bZ^\nu$. We denote by $\cB(\cA)$ the set of all translation-invariant
interactions $\Phi$ in $\cA$ of relatively short range, i.e.,
$|||\Psi|||:=\sum_{X\ni0}\|\Psi(X)\|/|X|<+\infty$, which is a real Banach space with
the norm $|||\Psi|||$. Let $\Phi\in\cB(\cA)$ and assume further that $\Phi$ is of
finite body, i.e., $N(\Phi):=\sup\{|X|:\Phi(X)\ne0\}<+\infty$ (weaker than the
assumption of finite range). Then $\Phi$ is automatically of short range, i.e.,
$\|\Phi\|:=\sum_{X\ni0}\|\Phi(X)\|<+\infty$. It is well known \cite{BR2,Is} that the
one-parameter automorphism group $\alpha_t^\Phi$ of $\cA$ is defined and all of the
$\alpha^\Phi$-KMS condition, the Gibbs condition and the variational principle for
states $\ffi\in\cS_\gamma(\cA)$ are equivalent. The pressure \eqref{F-1.5} and the
mean entropy \eqref{F-1.6}, the main ingredients in the variational principle, can be
defined in the van Hove limit of $\Lambda\to\infty$ (see \cite[p.\,12]{Is} or
\cite[p.\,287]{BR2}), but in our further discussions we may simply restrict to the
parallelepipeds $\Lambda=\{(k_1,\dots,k_\nu):1\le k_i\le n_i,\,1\le i\le\nu\}$ with
$\Lambda\to\infty$ meaning $n_i\to\infty$ for $1\le i\le\nu$.

A crucial point in the arbitrary dimensional setting is the following generalization
of Lemma \ref{L-2.1} given in \cite{AI} in high temperature regime with an inverse
temperature $\beta$.

\begin{lemma}\label{L-5.1}
Let $\Phi$ be given as above and $r(\Phi):=\{2\|\Phi\|(N(\Phi)-1)\}^{-1}$ {\rm(}meant
$+\infty$ if $N(\Phi)\le1${\rm)}. Assume that $0<\beta<2r(\Phi)$ and
$\ffi\in\cS_\gamma(\cA)$ satisfies the Gibbs condition for $\beta\Phi$
{\rm(}equivalently, the $\alpha^\Phi$-KMS condition at $-\beta${\rm)}. Then there are
constants $\lambda_\Lambda$ such that
$$
\lambda_\Lambda^{-1}\ffi_\Lambda\le\ffi_\Lambda^{\beta,G}
\le\lambda_\Lambda\ffi_\Lambda
$$
and
\begin{equation}\label{F-5.1}
\lim_{\Lambda\to\infty}{\log\lambda_\Lambda\over|\Lambda|}=0\,,
\end{equation}
where $\ffi_\Lambda^{\beta,G}$ is the local Gibbs state of $\cA_\Lambda$ for
$\beta\Phi$.
\end{lemma}

Even though a Gibbs state $\ffi\in\cS_\gamma(\cA)$ for $\beta\Phi$ is not necessarily
unique and constants $\lambda_\Lambda$ are depending on $\Lambda$, property
\eqref{F-5.1} is enough for us to show all the results in Section 2 (except the
differentiability assertions mentioned above) in the same way under the situation
where $\Phi$ is replaced by $\beta\Phi$ with $\beta$ as in Lemma \ref{L-5.1} and
$\cB_0(\cA)$ is replaced by $\cB(\cA)$. In particular, it was formerly observed in
\cite[p.\,710--711]{HP2} that for every $\omega\in\cS_\gamma(\cA)$ the mean relative
entropy \eqref{F-2.1} exists and furthermore $S_M(\omega,\ffi)=0$ if and only if
$\omega$ is a Gibbs state for $\beta\Phi$ too. In fact, the latter assertion is
immediate from the formula in Lemma \ref{L-2.3} due to the equivalence of the Gibbs
condition and the variational principle.

Next let $A\in\cA_\loc^\sa$ so that we may assume that $A\in\cA_{\Lambda_0}^\sa$ with
some parallelepiped $\Lambda_0\subset\bZ^\nu$ of the form mentioned above. Let $f$ be
a real continuous function on $[\lambda_{\min}(A),\lambda_{\max}(A)]$. For each
parallelepiped $\Lambda$ of the same form, we set
$$
s_\Lambda(A):={1\over|\Lambda|}\sum_{\Lambda_0+k\subset\Lambda}\gamma_k(A)
$$
and
$$
Z_\ffi(\Lambda,A,f)
:=\Tr\exp\bigl(\log D(\ffi_\Lambda)-|\Lambda|f(s_\Lambda(A))\bigr)\,.
$$
Then Lemmas \ref{L-3.1} and \ref{L-3.2} hold true in the same way as before. Moreover,
the proof of Lemma \ref{L-3.3} can easily be carried out in the present framework with
slight modifications, for example, with replacing the uniform boundedness of surface
energies by the asymptotic property
$$
{1\over|\Lambda|}
\sum\{\|\Phi(X)\|:X\cap\Lambda\ne\emptyset,\,X\cap\Lambda^c\ne\emptyset\}
\longrightarrow0
$$
as $\Lambda\to\infty$ of parallelepipeds $\Lambda$. This property holds in general for
translation-invariant interactions of short range.

Finally, we can prove the existence of the functional free
energy density
$$
p_\ffi(A,f):=\lim_{\Lambda\to\infty}{1\over|\Lambda|}\log Z_\ffi(\Lambda,A,f)
$$
and its variational expressions in the same way as in Theorem \ref{T-3.4}. A key point
in proving this is that the result for the product state case in \cite{PRV} (or
\cite{OP}) used in the proof of Theorem \ref{T-3.4} can be applied as well since the
dimension of the integer lattice is irrelevant in the situation of product/symmetric
states. In this way, all the proofs in Section 3 of one dimension can easily be
adapted to the present framework by using Lemma \ref{L-5.1} and the property of short
range for $\Phi$, and the condition of finite range is not necessary.

\section*{Acknowledgements}
The authors are grateful to an anonymous referee for his comments that are very helpful
in improving the final version of the paper.

\end{document}